\documentclass[preprint,3p,12pt]{elsarticle}

\usepackage{amssymb}
\usepackage{amsthm}

\usepackage{array}
\usepackage{import}
\usepackage{tikz}
\usetikzlibrary{arrows.meta}
\usepackage{calc}

\usepackage{ctable}
\usepackage{multirow}

\usepackage{xcolor}
\definecolor{light-gray}{gray}{0.50}                    
\definecolor{black}{rgb}{0,0,0}                 
\definecolor{darkgreen}{rgb}{0.0 0.5 0.0}       
\definecolor{lightblue}{rgb}{0.0 0.5 1.0}       
\definecolor{dblue}{rgb}{0.400 0.400 0.800}     
\definecolor{lgray}{rgb}{0.800 0.800 0.800}     
\definecolor{ddblue}{rgb}{0.1484 0.2266 0.4258} 
\definecolor{RIVASgreen}{rgb}{0.2 0.6 0.4}      
\definecolor{dyellow}{rgb}{1 0.63 0}

\usepackage{graphicx}
\usepackage{float}
\usepackage{amsfonts}
\usepackage{commath}
\usepackage{url}

\newcolumntype{L}[1]{>{\raggedright\let\newline\\\arraybackslash\hspace{0pt}}m{#1}}
\newcolumntype{C}[1]{>{\centering\let\newline\\\arraybackslash\hspace{0pt}}m{#1}}
\newcolumntype{R}[1]{>{\raggedleft\let\newline\\\arraybackslash\hspace{0pt}}m{#1}}

\usepackage[most]{tcolorbox}
\tcbuselibrary{theorems}

\newtcbtheorem[number within=section]{remark}{Remark}%
{enhanced,
 breakable,
 colback=black!10,
 colframe=black!10,
 boxrule=0pt,
 sharp corners,
 left=5pt,
 right=5pt,
 top=5pt,
 bottom=5pt,
 fonttitle=\bfseries,
 coltitle=black
}{remark}

\newcommand{\nrm}[1]{\left \lVert #1 \right \rVert}

\definecolor{blue2}{RGB}{25, 132, 197}
\definecolor{red2}{RGB}{194, 55, 40}
\definecolor{gray2}{RGB}{102, 102, 102}
\definecolor{darkgreen}{rgb}{0.0 0.5 0.0}       

\DeclareRobustCommand\full  {\tikz[baseline=-0.6ex]\draw[very thick] (0,0)--(0.5,0);}
\DeclareRobustCommand\dotted{\tikz[baseline=-0.6ex]\draw[ultra thick,dotted] (0,0)--(0.46,0);}
\DeclareRobustCommand\dashed{\tikz[baseline=-0.6ex]\draw[very thick,dashed] (0,0)--(0.54,0);}

\journal{\textcolor{gray}{To be determined.}}

\begin{document}

\begin{frontmatter}



\title{An efficient finite element formulation for Newtonian noise analysis\tnotemark[1]}

\tnotetext[1]{{\itshape Published version:}
P.~Reumers, X.~Kuci, S.~Fran\c{c}ois and G.~Degrande.
\newblock An efficient finite element formulation for Newtonian noise analysis.
\newblock {\em Computer Physics Communications},
329, 110390, 2026. \\
\url{https://doi.org/10.1016/j.cpc.2026.110390}}

\author[KULeuven]{P. Reumers\fnref{fn1}} 
\author[KULeuven]{X. Kuci\corref{cor1}\fnref{fn1}} \ead{georgia.kuci@kuleuven.be}
\author[KULeuven]{S. Fran\c{c}ois} 
\author[KULeuven]{G. Degrande} 

\affiliation[KULeuven]{organization={KU Leuven, Department of Civil Engineering, 
Structural Mechanics Section},
            addressline={Kasteelpark Arenberg 40}, 
            city={Leuven},
            postcode={B-3001}, 
            country={Belgium}}
            
\cortext[cor1]{Corresponding author}
\fntext[fn1]{These authors should be regarded as joint first authors.}

\begin{abstract}
The Einstein Telescope is a third-generation underground gravitational wave 
observatory designed to achieve an unprecedented sensitivity down to 3 Hz. 
Waves propagating in the soil due to anthropogenic or natural vibration sources
generate density fluctuations which cause gravitational attraction, resulting in 
motion of the mirrors of the laser interferometer known as Newtonian noise.
The latter is computed by integrating density fluctuations due to seismic wave 
fields over the soil domain surrounding the test mass.
\par
A finite element formulation is presented, which evaluates the total Newtonian
noise, as well as the bulk and surface contributions, from a seismic wave field
defined on a finite element mesh, using Gaussian quadrature.
Linear and quadratic tetrahedral (4-node and 10-node) and brick (8-node and 20-node) 
finite elements are supported.
The approach computes the total, bulk, and surface contributions. The corresponding volume and surface integrals are expressed in terms of finite element coupling matrices that depend only on the geometry and material properties. 
This allows the Newtonian noise to be evaluated efficiently for different seismic wave fields without recomputing the integrals.
\par
The formulation is verified for plane P- and S-waves propagating in an elastic
homogeneous full space with a mirror suspended in a spherical cavity, assuming 
that the wavelength is much larger than the radius of the cavity, so that 
wave scattering can be ignored.
Excellent agreement with analytical solutions is obtained.
Similar good agreement is reported for the Newtonian noise on a test mass suspended 
at a finite distance above the free surface of a homogeneous elastic halfspace 
in which a Rayleigh wave propagates.
\par
The methodology has been implemented in the ANNA Newtonian Noise Analysis toolbox
which runs in the MATLAB programming and numeric computing platform and is 
compatible with the open-source GNU Octave Scientific Programming Language;
a Python version is also available. The proposed finite element framework provides 
a physically consistent and computationally efficient approach for computing 
gravitational-seismic coupling in heterogeneous media.
\end{abstract}

%

\end{frontmatter}


\newpage
\section{Introduction}
\label{sec:intro}

The Einstein Telescope is a third-generation gravitational wave detector that is 
planned to be constructed in the coming decade \cite{ET-DR-2020}. 
It consists of an underground triangular laser interferometer and is designed to 
observe gravitational waves \textendash~ripples in the 4D space-time continuum 
\textendash~due to e.g.~binary black hole or binary neutron star mergers. 
The mirrors of the laser interferometer are installed in suspension towers placed
inside large caverns at the corner points of the interferometer.
\par
To observe gravitational waves caused by mergers at a larger distance or with 
a larger mass, the sensitivity and operating frequency range (above 10 Hz) of 
second-generation gravitational wave detectors such as LIGO and Virgo are 
insufficient \cite{ET-DR-2020}.
In order to improve the sensitivity by one order of magnitude and to reduce the 
operating frequency to 3 Hz, the Einstein Telescope is preferably constructed 
underground in a seismically quiet region~\cite{aman20a} so that disturbance 
by anthropogenic vibration sources such as road and railway traffic, wind 
turbines, quarries, industry and construction activities is maximally reduced.
\par
Seismic Newtonian noise dominates at low frequencies: seismic waves propagating 
in the soil generate density fluctuations, which result in gravitational attraction 
and corresponding motion of the mirrors of the laser interferometer. 
Since these gravitational forces cannot be shielded, the only mitigation measure
is to estimate Newtonian noise from seismic measurements through wave field 
reconstruction and to subtract it from the interferometer data~\cite{drig12a}.
\par
Analytical solutions for Newtonian noise are available for simplified problems
involving plane harmonic P- or S-waves propagating in a homogeneous linear elastic 
full space~\cite{harm19a}, assuming that the wavelength is much larger than 
the dimensions of the cavity so that wave scattering can be neglected.
Harms~\cite{harm19a} also derived an analytical solution for the Newtonian
noise on a mirror suspended at a finite distance above the free surface of a 
homogeneous linear elastic halfspace, excited by a harmonic Rayleigh wave.
In practice, however, the topology and geology of a site are more complicated,
involving hills, valleys, stratification due to sedimentation processes, as
well as folding of soil and rock layers.
Seismic wave fields are caused by a variety of anthropogenic vibration sources 
and affected by the soil's topology and layering, as well as the interferometer's 
caverns, tunnels, and shafts.
As a result, numerical methods are required to compute the seismic wave field
and the corresponding Newtonian noise. 
\par
Several methods are available to compute a three-dimensional (3D) seismic wave 
field due to anthropogenic vibration sources, accounting for dynamic soil-structure
interaction.
A direct stiffness formulation or thin layer method \cite{kaus81a}, as 
implemented in the ElastoDynamics Toolbox for MATLAB \cite{ij-cgs-sche-09a}, 
is used to model surface waves or forced vibration due to harmonic or transient 
excitation in horizontally layered media.
Vibration due to road and railway traffic can be computed with the TRAFFIC 
toolbox in MATLAB, which couples a semi-analytical road or track model to
a two-and-a-half (2.5D) dimensional boundary element (BE) model of the layered soil
\cite{ij-jsv-lomb-06a,ij-jsv-lomb-09a,bwm-2012-10}; 
alternatively the MOTIV~\cite{shen04a,ntot17a,ntot19a} and MEFISSTO~\cite{jean15a} 
packages can be used. 
3D finite element formulations equipped with infinite elements, absorbing boundary 
conditions or perfectly matched layers
\cite{yang08a,yang09a,ij-ijnme-fran-12a,ande14a}, 
coupled finite element-boundary element formulations
\cite{ij-cmame-fran-10a,jean15a,ij-jsv-mpap-18a} 
and spectral element models such as SPECFEM3D and SPEED \cite{mazz13a} can 
be employed to solve more complicated dynamic soil-structure interaction problems,
such as vibration caused by trains running on bridges or by wind turbines and
wave scattering by the interferometer's caverns, tunnels and shafts.
\par
Recent numerical studies have evaluated Newtonian 
noise by first simulating a seismic wave field and then performing a direct numerical evaluation of the resulting Newtonian noise volume integral.
For instance, in \cite{bade22a} the direct stiffness method was utilized to solve the elastodynamic equations for horizontally layered media, subsequently computing the Newtonian noise via Gaussian quadrature.
Similarly, in \cite{schil26a}, spectral element simulations were employed to model the seismic wave field. The corresponding Newtonian noise was subsequently evaluated through a dipole-based volume integral.
In these approaches, the Newtonian noise integral is treated as a post-processing step dependent on a specific seismic realization, requiring a full re-evaluation for each seismic wave field.
Assuming that the seismic wave field is computed on a 3D finite element mesh sufficiently refined to accurately resolve wave propagation, this paper presents an efficient finite element formulation for evaluating Newtonian noise by numerically integrating its total, bulk, and surface contributions using Gaussian quadrature. The formulation yields geometry-dependent matrices that are assembled once and subsequently applied to any seismic wave field. The resulting framework enables computationally efficient Newtonian noise analysis in heterogeneous three-dimensional media while maintaining the numerical accuracy required to resolve gravitational fluctuations in the vicinity of the test mass.
The finite element formulation is verified by means of the 
aforementioned analytical formulations, considering the Newtonian noise 
on a mirror suspended in a spherical cavity in a homogeneous linear elastic 
full space, excited by plane P- or S-waves~\cite{harm19a}, as well as a mirror 
suspended at a finite distance above the free surface of a homogeneous linear 
elastic halfspace, excited by a Rayleigh wave~\cite{harm19a}.
\par
The paper is organized as follows.
Section~\ref{sec:theory} presents expressions for the total derivative of the 
gravitational potential, as well as total Newtonian noise which is decomposed 
in a bulk and surface contribution.
Section~\ref{sec:fe} develops a finite element formulation to evaluate the 
Newtonian noise for a seismic wave field defined on a finite element mesh.
Expressions are derived for finite element matrices that, when multiplied
with the seismic displacement vector, result in the total Newtonian noise,
as well as the bulk and surface contribution.
Section~\ref{sec:verification} presents the verification of the proposed finite
element formulation by means of problems for which an analytical solution is 
available. 
Section~\ref{sec:conclusion} summarizes the conclusions of the paper.

\section{Theoretical background}
\label{sec:theory}

Seismic Newtonian noise is the undesired motion (acceleration) caused by
gravitational attraction of the mirror of the laser interferometer due to 
density fluctuations in the soil, that are caused by seismic wave fields.

\subsection{Gravitational potential}
\label{sec:gp}

The gravitational force $F$ between two point masses $M$ and $m$ at a distance $r$ 
is equal to:
\begin{eqnarray}
   F &=& G \frac{Mm}{r^{2}} \,,
   \label{eq:gf1}
\end{eqnarray}
where $G \approx 6.6743\times 10^{-11} \,\mbox{m}^{3}/\mbox{kg}/\mbox{s}^{2}$ 
is the gravitational constant.
The gravitational potential $\phi(r)$ [Nm/kg] is defined as the work 
per unit mass $m$ required to move this mass $m$
from infinity (where $\phi=0$) to a distance 
$r$ from the mass $M$:
\begin{eqnarray}
   \phi(r) &=& 
   \frac{1}{m} \int_{+\infty}^{r} G \frac{Mm}{x^{2}} \, \mathrm{d}x
   \, = \,
   - G \frac{M}{r} \,.
   \label{eq:gp1}
\end{eqnarray}
The value of $\phi(r)$ is negative as work is required to pull objects away 
from large masses.
\par 
In equation (\ref{eq:gp1}), the mass $m$ can be interpreted as the mass of a 
mirror of the laser interferometer at position ${\bf x}_{0}$. 
Furthermore, the point mass $M$ is replaced by a set of particles that at time 
$t=0$ are contained in a volume $\Omega({\bf X})$, where ${\bf X}$ are the 
coordinates in the reference configuration.
When a dynamic excitation causes a seismic wave field, this set of particles 
moves and, at time $t$, is contained in a volume $\Omega({\bf x},t)$ in the 
actual configuration ${\bf x}$. 
The gravitational potential $\phi({\bf x}_{0},t)$ at the position ${\bf x}_{0}$ 
of the mirror with mass $m$, or the work per unit mass $m$, is then expressed as:
\begin{eqnarray}
   \phi({\bf x}_{0},t) &=&
   - G \int_{\Omega({\bf x},t)} 
   \frac{\rho({\bf x},t)}{\nrm{{\bf x}-{\bf x}_{0}}} 
   \, \mathrm{d}v \,,
   \label{eq:gp2}
\end{eqnarray}
where the density $\rho({\bf x},t)$ of the material is written 
in terms of the actual configuration ${\bf x}$ and the time $t$.
\begin{figure}[!htb]
	\centering
	\includegraphics[scale=1]{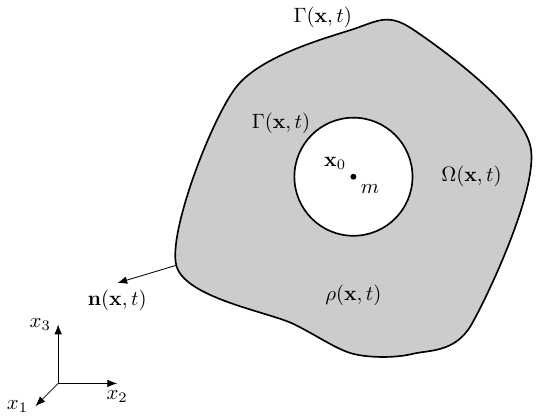}
    \caption{Point mass $m$ at a position ${\bf x}_{0}$ and 
     volume $\Omega({\bf x},t)$ with density $\rho({\bf x},t)$.}
\end{figure}

\subsection{Total derivative of the gravitational potential}

A seismic wave field ${\bf u}({\bf x},t)$ causes density variations and, 
consequently, a variation of the gravitational potential $\phi({\bf x}_{0},t)$.
This variation is expressed as a total derivative with respect to time:
\begin{eqnarray}
   \frac{D \phi({\bf x}_{0},t)}{Dt} &=&
   - G \frac{D}{Dt}
   \int_{\Omega({\bf x},t)} 
    \frac{\rho({\bf x},t)}{\nrm{{\bf x} - {\bf x}_{0}}}
   \, \mathrm{d}v \,,
   \label{eq:tdgp1}   
\end{eqnarray}
where it is understood that both the integrand as well as the volume
$\Omega({\bf x},t)$ depend on time~$t$. 
The density $\rho({\bf x},t)$ depends on time $t$ due to the time dependence
of the volumetric strain which results from the seismic wave field 
${\bf u}({\bf x},t)$, as expressed in the conservation of mass.
\par
The total derivative in equation (\ref{eq:tdgp1}) is elaborated by means of Reynolds' transport theorem (\ref{app:rtt}), assuming that the density $\rho(\mathbf{x},t)$ and the seismic wave field $\mathbf{u}(\mathbf{x},t)$ are continuously differentiable.
Referring to equation (\ref{eq:rtt1}), the scalar function $g({\bf x},t)$ in the 
present case is equal to $- G \rho({\bf x},t) / \nrm{ {\bf x} - {\bf x}_{0} }$, 
and the total derivative of the gravitational potential $\phi({\bf x}_{0},t)$ with 
respect to time is equal to:
\begin{eqnarray}
   \frac{D \phi({\bf x}_{0},t)}{Dt} &=&
   - G  \int_{\Omega({\bf x},t)}
   \left[
   \frac{\partial}{\partial t} 
   \left( \frac{\rho({\bf x},t)}{\nrm{{\bf x} - {\bf x}_{0}}} \right)   
   +
   \nabla \cdot  
   \left(
   \frac{\rho({\bf x},t) {\bf v}({\bf x},t)}{\nrm{{\bf x} - {\bf x}_{0}}}
   \right)
   \right]
   \, \mathrm{d}v
   \nonumber \\ &=&
   - G  \int_{\Omega({\bf x},t)}
   \left[
   \frac{\partial}{\partial t} 
   \left( \frac{\rho({\bf x},t)}{\nrm{{\bf x} - {\bf x}_{0}}} \right)   
   + \frac{1}{\nrm{{\bf x} - {\bf x}_{0}}}
   \nabla \cdot
   \left( \rho({\bf x},t) {\bf v}({\bf x},t)
   \right)
   \right. \nonumber \\ && \left.
   + \rho({\bf x},t) {\bf v}({\bf x},t) \cdot
   \nabla \frac{1}{\nrm{{\bf x} - {\bf x}_{0}}}
   \right]
   \, \mathrm{d}v
   \nonumber \\ &=&
   - G  \int_{\Omega({\bf x},t)}
   \left[
   \frac{1}{\nrm{{\bf x} - {\bf x}_{0}}}   
   \left(
   \frac{\partial \rho({\bf x},t)}{\partial t} +
   {\bf v}({\bf x},t) \cdot \nabla \rho({\bf x},t) +
   \rho({\bf x},t) \nabla \cdot {\bf v}({\bf x},t)
   \right)
   \right. \nonumber \\ && \left.
   + \rho({\bf x},t) {\bf v}({\bf x},t) \cdot
   \nabla \, \frac{1}{\nrm{{\bf x} - {\bf x}_{0}}}
   \right]
   \, \mathrm{d}v  \,,
   \label{eq:tdgp2}  
\end{eqnarray}
where $\nabla$ denotes the gradient and $\nabla \cdot$ the divergence.
According to the conservation of mass (\ref{app:com}), the term between
round brackets in equation (\ref{eq:tdgp2}) is equal to zero. 
The total derivative of the gravitational potential hence reduces to:
\begin{eqnarray}
   \frac{D \phi({\bf x}_{0},t)}{Dt} &=&
   - G  \int_{\Omega({\bf x},t)}
   \rho({\bf x},t) {\bf v}({\bf x},t) \cdot
   \nabla \frac{1}{\nrm{{\bf x} - {\bf x}_{0}}}
   \, \mathrm{d}v \,.
   \label{eq:tdgp3}
\end{eqnarray}
An auxiliary function $\boldsymbol{\chi}({\bf x},{\bf x}_{0})$ is now introduced:
\begin{eqnarray}
   \boldsymbol{\chi}({\bf x},{\bf x}_{0}) &=&
   - \nabla     \frac{1}{\nrm{{\bf x} - {\bf x}_{0}}} \, = \,
   + \nabla_{0} \frac{1}{\nrm{{\bf x} - {\bf x}_{0}}} \, = \,
   \frac{{\bf x} - {\bf x}_{0}}{\nrm{{\bf x} - {\bf x}_{0}}^{3}} \,,
 \end{eqnarray}
where $\nabla$ and $\nabla_{0}$ are the gradient operators with respect to the
coordinates ${\bf x}$ and ${\bf x}_{0}$, respectively.
The total derivative of the gravitational potential is finally written as:
\begin{eqnarray}
   \frac{D \phi({\bf x}_{0},t)}{Dt} &=&
   + G  \int_{\Omega({\bf x},t)}
   \rho({\bf x},t) {\bf v}({\bf x},t) \cdot
   \boldsymbol{\chi}({\bf x},{\bf x}_{0})
   \, \mathrm{d}v \,.
   \label{eq:tdgp4}
\end{eqnarray}
In equation (\ref{eq:tdgp4}), the volume $\Omega({\bf x},t)$ can also be replaced 
by a fixed control volume $\bar{\Omega}$ (\ref{app:rtt}):
\begin{eqnarray}
   \frac{D \phi({\bf x}_{0},t)}{Dt} &=&
   + G  \int_{\bar{\Omega}}
   \rho({\bf x},t) {\bf v}({\bf x},t) \cdot
   \boldsymbol{\chi}({\bf x},{\bf x}_{0})
   \, \mathrm{d}v  \,.
   \label{eq:tdgp5}
\end{eqnarray}
\par
An alternative formulation of the total derivative of the gravitational 
potential is obtained from Reynolds' transport theorem (\ref{eq:rtt3d}):
\begin{eqnarray}
   \frac{D \phi({\bf x}_{0},t)}{Dt} &=&
   - G \int_{\Omega({\bf x},t)}
   \frac{\partial}{\partial t} 
   \left( \frac{\rho({\bf x},t)}{\nrm{{\bf x} - {\bf x}_{0}}} \right)   
   \, \mathrm{d}v
   - G \int_{\Gamma({\bf x},t)}
   \frac{\rho({\bf x},t)}{\nrm{{\bf x} - {\bf x}_{0}}}
   {\bf v}({\bf x},t) \cdot {\bf n}({\bf x},t)
   \, \mathrm{d}a   
   \label{eq:tdgp6a}  
   \nonumber \\ &=& 
   - G \int_{\Omega({\bf x},t)}
   \frac{1}{\nrm{{\bf x} - {\bf x}_{0}}}
   \frac{\partial \rho({\bf x},t)}{\partial t} 
   \, \mathrm{d}v
   - G \int_{\Gamma({\bf x},t)}
   \frac{\rho({\bf x},t)}{\nrm{{\bf x} - {\bf x}_{0}}}
   {\bf v}({\bf x},t) \cdot {\bf n}({\bf x},t)
   \, \mathrm{d}a   
   \label{eq:tdgp6b}  
   \nonumber \\ &=&
   + G \int_{\Omega({\bf x},t)}
   \frac{1}{\nrm{{\bf x} - {\bf x}_{0}}}
   \nabla \cdot 
   \left( \rho({\bf x},t) {\bf v}({\bf x},t) \right)
   \, \mathrm{d}v
   \nonumber \\ &&
   - G \int_{\Gamma({\bf x},t)}
   \frac{\rho({\bf x},t)}{\nrm{{\bf x} - {\bf x}_{0}}}
   {\bf v}({\bf x},t) \cdot {\bf n}({\bf x},t)
   \, \mathrm{d}a \,,  
   \label{eq:tdgp6c}  
\end{eqnarray}
differentiating between a bulk and surface contribution.
In equation (\ref{eq:tdgp6c}), the volume $\Omega({\bf x},t)$ with boundary 
$\Gamma({\bf x},t)$ can also be replaced by a fixed control volume 
$\bar{\Omega}$ with boundary $\bar{\Gamma}$ (\ref{app:rtt}):
\begin{eqnarray}
   \frac{D \phi({\bf x}_{0},t)}{Dt} =
   + G \int_{\bar{\Omega}}
   \frac{1}{\nrm{{\bf x} - {\bf x}_{0}}}
   \nabla \cdot 
   \left( \rho({\bf x},t) {\bf v}({\bf x},t) \right)
   \, \mathrm{d}v
   - G \int_{\bar{\Gamma}}
   \frac{\rho({\bf x},t)}{\nrm{{\bf x} - {\bf x}_{0}}}
   {\bf v}({\bf x},t) \cdot {\bf n}({\bf x},t)
   \, \mathrm{d}a \,.  
   \label{eq:tdgp7}
\end{eqnarray}

\subsection{Variation of the gravitational potential}

As discussed in Appendix~A, Reynolds' transport theorem provides a formal
expression of both the material derivative and the corresponding variation
of a quantity observed when following a set of particles.
In a rate-independent form where the total derivative of the gravitational 
potential with respect to time is replaced by a variation, and the velocity 
vector is replaced by the displacement vector, the variation of the
gravitational potential becomes:
\begin{eqnarray}
   \delta \phi({\bf x}_{0},t) &=&
   + G \int_{\bar{\Omega}}
   \rho({\bf x},t) {\bf u}({\bf x},t) \cdot
   \boldsymbol{\chi}({\bf x},{\bf x}_{0})
   \, \mathrm{d}v \,.
  \label{eq:vgp1}
\end{eqnarray}
This equation is equivalent to equation (43) derived by Harms \cite{harm19a}.
\par
Differentiating between the bulk and the surface contribution according
to equation (\ref{eq:tdgp7}), the equivalent expression for the variation 
of the gravitational potential becomes:
\begin{eqnarray}
   \delta \phi({\bf x}_{0},t) =
   + G \int_{\bar{\Omega}}
   \frac{1}{\nrm{{\bf x} - {\bf x}_{0}}}
   \nabla \cdot \left( \rho({\bf x},t) {\bf u}({\bf x},t) \right)
   \, \mathrm{d}v
   - G \int_{\bar{\Gamma}} 
   \frac{\rho({\bf x},t)}{\nrm{{\bf x} - {\bf x}_{0}}}
   {\bf u}({\bf x},t) \cdot {\bf n}({\bf x},t)
   \, \mathrm{d}a \,.  
   \label{eq:vgp2}  
\end{eqnarray}
The first integral corresponds to the bulk contribution, while the second
integral is the surface contribution.
These expressions are equivalent to equations (46) and (47) derived by
Harms \cite{harm19a} for a homogeneous medium.

\subsection{Newtonian noise}

The total Newtonian noise or the perturbation~$\delta{\bf a}_{\mathrm{t}}({\bf x}_{0},t)$ 
[m/s$^{2}$] is obtained as the negative gradient of the variation 
of the gravitational potential $\delta \phi({\bf x}_{0},t)$ with respect to the 
mirror position ${\bf x}_{0}$:
\begin{eqnarray}
   \delta{\bf a}_{\mathrm{t}}({\bf x}_{0},t) &=&
   - \nabla_{0} \delta \phi({\bf x}_{0},t) \,.
\end{eqnarray}
Based on equation (\ref{eq:tdgp7}), the following expression is obtained for the
total Newtonian noise: 
\begin{eqnarray}
   \delta{\bf a}_{\mathrm{t}}({\bf x}_{0},t) &=&
   - G \int_{\bar{\Omega}}
   \rho({\bf x},t) \nabla_{0} \left( {\bf u}({\bf x},t) \cdot
   \boldsymbol{\chi}({\bf x},{\bf x}_{0}) \right)
   \, \mathrm{d}v 
   \nonumber \\ &=&
   - G \int_{\bar{\Omega}} 
   \rho({\bf x},t) 
   \nabla_{0} \boldsymbol{\chi}({\bf x},{\bf x}_{0}) \cdot {\bf u}({\bf x},t)
   \, \mathrm{d}v 
   \nonumber \\ &=&
   + G \int_{\bar{\Omega}} 
   \frac{\rho({\bf x},t)}{\nrm{{\bf x} - {\bf x}_{0}}^{3}}
   ({\bf I}_{3} - 3 {\bf e}_{\mathrm{r}} \otimes {\bf e}_{\mathrm{r}})
   {\bf u}({\bf x},t)
   \, \mathrm{d}v \,,
   \label{eq:nnt1}  
\end{eqnarray}
where ${\bf I}_{3}$ is the 3 by 3 identity matrix and the vector 
${\bf e}_{\mathrm{r}}$ is defined as:
\begin{eqnarray}
   {\bf e}_{\mathrm{r}} &=& 
   \frac{{\bf x} - {\bf x}_{0}}{\nrm{{\bf x} - {\bf x}_{0}}} \,.
   \label{eq:er}
\end{eqnarray}
In the frequency domain, equation~\eqref{eq:nnt1} becomes:
\begin{eqnarray}
   \delta\hat{\bf a}_{\mathrm{t}}({\bf x}_{0},\omega) &=&
  + G \int_{\bar{\Omega}}
  \frac{\rho({\bf x})}{\nrm{{\bf x} - {\bf x}_{0}}^{3}}
  ({\bf I}_{3} - 3 {\bf e}_{\mathrm{r}} \otimes {\bf e}_{\mathrm{r}}) 
  \hat{\bf u}({\bf x},\omega)
  \, \mathrm{d}v \,,
  \label{eq:nntf1}
\end{eqnarray}
where a hat on a variable denotes its representation in the frequency domain.
\par
Based on equation (\ref{eq:tdgp7}), the bulk and surface contribution to the total
Newtonian noise can be separated: 
\begin{eqnarray}
   \delta{\bf a}_{\mathrm{t}}({\bf x}_{0},t) &=&
   \delta{\bf a}_{\mathrm{b}}({\bf x}_{0},t) +
   \delta{\bf a}_{\mathrm{s}}({\bf x}_{0},t) \,,
   \label{eq:nnt2}
\end{eqnarray}
where the bulk contribution 
$\delta{\bf a}_{\mathrm{b}}({\bf x}_{0},t)$ is equal to:
\begin{eqnarray}
   \delta{\bf a}_{\mathrm{b}}({\bf x}_{0},t) &=&
   - G \int_{\bar{\Omega}}
   \boldsymbol{\chi}({\bf x},{\bf x}_{0})
   \nabla \cdot \left( \rho({\bf x},t) {\bf u}({\bf x},t) \right)
   \, \mathrm{d}v
   \label{eq:nnb1}
\end{eqnarray}
and the surface contribution 
$\delta{\bf a}_{\mathrm{s}}({\bf x}_{0},t)$ is equal to:
\begin{eqnarray}
   \delta{\bf a}_{\mathrm{s}}({\bf x}_{0},t) &=&
   + G \int_{\bar{\Gamma}}
   \boldsymbol{\chi}({\bf x},{\bf x}_{0})
   \rho({\bf x},t)
   {\bf u}({\bf x},t) \cdot {\bf n}({\bf x},t)
   \, \mathrm{d}a \,.  
   \label{eq:nns1}  
\end{eqnarray}
In the frequency domain, the bulk contribution 
$\delta\hat{\bf a}_{\mathrm{b}}({\bf x}_{0},\omega)$ is equal to:
\begin{eqnarray}
  \delta\hat{\bf a}_{\mathrm{b}}({\bf x}_{0},\omega) &=&
  - G \int_{\bar{\Omega}}
  \boldsymbol{\chi}({\bf x},{\bf x}_{0})
  \nabla \cdot \left( \rho({\bf x}) \hat{\bf u}({\bf x},\omega) \right)
  \, \mathrm{d}v \,,
  \label{eq:nnbf1}
\end{eqnarray}
while the surface contribution
$\delta\hat{\bf a}_{\mathrm{s}}({\bf x}_{0},\omega)$ is equal to:
\begin{eqnarray}
  \delta\hat{\bf a}_{\mathrm{s}}({\bf x}_{0},\omega) &=&
  + G \int_{\bar{\Gamma}}
  \boldsymbol{\chi}({\bf x},{\bf x}_{0})
  \rho({\bf x})
  \hat{\bf u}({\bf x},\omega) \cdot {\bf n}({\bf x})
  \, \mathrm{d}a \,.
  \label{eq:nnsf1}
\end{eqnarray}

\section{Finite element formulation}
\label{sec:fe}

\subsection{Total Newtonian noise}

In order to evaluate the volume integral in equation~\eqref{eq:nntf1}, the control 
volume $\bar{\Omega}$ is discretized into $n_{e}$ finite elements (FE) with volume 
$\bar{\Omega}^{e}$ and constant density $\rho_{e}$, resulting in the following sum:
\begin{eqnarray}
   \delta\hat{\bf a}_{\mathrm{t}}({\bf x}_{0},\omega) & \simeq &
   \sum\limits_{e=1}^{n_{e}} 
   G \int_{\bar{\Omega}^{e}} 
   \frac{\rho_{e}}{\nrm{{\bf x}-{\bf x}_{0}}^{3}}
   \left( {\bf I}_{3} - 
         3{\bf e}_{\mathrm{r}} \otimes {\bf e}_{\mathrm{r}} \right)
   \hat{\bf u}^{e}({\bf x},\omega)
   \, \mathrm{d}v \,,
   \label{eq:fe-nnt-1}
\end{eqnarray}
where $\hat{\bf u}^{e}({\bf x},\omega)$ are the element displacements.
\par
The element displacements $\hat{\bf u}^{e}(\boldsymbol{\xi},\omega)$ in the 
local coordinate system are approximated using the shape functions 
$N^{e}_{k}(\boldsymbol{\xi})$ 
\begin{eqnarray}
	\hat{\bf u}^{e}(\boldsymbol{\xi},\omega) & \simeq &
	\sum\limits_{k=1}^{n}
	{\bf N}^{e}_{k}(\boldsymbol{\xi}) \underline{\hat{\bf u}}^{e}_{k} \, = \,
	{\bf N}^{e}(\boldsymbol{\xi}) \underline{\hat{\bf u}}^{e} \,,
	\label{eq:fe-ue}
\end{eqnarray}
where 
$\underline{\hat{\bf u}}^{e}= \{
{\underline{\hat{\bf u}}^{e}_{1}}^\mathrm{T}, 
{\underline{\hat{\bf u}}^{e}_{2}}^\mathrm{T},\dots,
{\underline{\hat{\bf u}}^{e}_n}^\mathrm{T}\}^\mathrm{T}$ 
is a column vector containing the nodal values of 
$\hat{\bf u}^{e}(\boldsymbol{\xi},\omega)$ and 
${\bf N}^{e}(\boldsymbol{\xi})= \left[
{\bf N}^{e}_{1}(\boldsymbol{\xi}), 
{\bf N}^{e}_{2}(\boldsymbol{\xi}), \dots, 
{\bf N}^{e}_n(\boldsymbol{\xi}) \right]$.  
Within each element, an isoparametric transformation 
is employed to transform
the global coordinate ${\bf x}=\{ x_{1},x_{2},x_{3} \}^{\mathrm{T}}$ 
to a local (element) coordinate 
$\boldsymbol{\xi}=\{ \xi_{1},\xi_{2},\xi_{3} \}^{\mathrm{T}}$ 
using shape functions $N^{e}_{k}(\boldsymbol{\xi})$: 
\begin{eqnarray}
	{\bf x}(\boldsymbol{\xi}) &=&
	\sum\limits_{k=1}^{n}
	\begin{bmatrix}
		N^{e}_{k}(\boldsymbol{\xi}) & 0 & 0 \\
		0 & N^{e}_{k}(\boldsymbol{\xi}) & 0 \\
		0 & 0 & N^{e}_{k}(\boldsymbol{\xi}) 
	\end{bmatrix}
	{\bf x}_{k}
    \, = \,
	\sum\limits_{k=1}^{n} {\bf N}^{e}_{k}(\boldsymbol{\xi}) {\bf x}_{k} \,,
	\label{eq:fe-xe}
\end{eqnarray}
where $n$ is the number of nodes in the element and ${\bf x}_{k}$ is the global
coordinate of node $k$. 
The integral in equation~\eqref{eq:fe-nnt-1} over the volume $\bar{\Omega}^{e}$ 
of each element is subsequently transformed to the local coordinate 
$\boldsymbol{\xi}$:
\begin{eqnarray}
    \mathrm{d}v &=&
	\det({\bf J}^{e}(\boldsymbol{\xi}))
	\mathrm{d}\xi_{1} \, \mathrm{d}\xi_{2} \, \mathrm{d}\xi_{3} \,,
	\label{eq:fe-dv}
\end{eqnarray}
where the Jacobian ${\bf J}^{e}(\boldsymbol{\xi})$ is defined as:
\begin{eqnarray}
    {\bf J}^{e}(\boldsymbol{\xi}) &=&
    \begin{bmatrix}
        \frac{\partial x_{1}}{\partial \xi_{1}} & 
        \frac{\partial x_{2}}{\partial \xi_{1}} & 
        \frac{\partial x_{3}}{\partial \xi_{1}} \\
        \frac{\partial x_{1}}{\partial \xi_{2}} & 
        \frac{\partial x_{2}}{\partial \xi_{2}} & 
        \frac{\partial x_{3}}{\partial \xi_{2}} \\
        \frac{\partial x_{1}}{\partial \xi_{3}} & 
        \frac{\partial x_{2}}{\partial \xi_{3}} & 
        \frac{\partial x_{3}}{\partial \xi_{3}}
    \end{bmatrix}
    \, = \,
    \sum\limits_{k=1}^{n}
    \begin{bmatrix}
        \frac{\partial N^{e}_{k}(\boldsymbol{\xi})}{\partial \xi_{1}}x_{k1} &
        \frac{\partial N^{e}_{k}(\boldsymbol{\xi})}{\partial \xi_{1}}x_{k2} &
        \frac{\partial N^{e}_{k}(\boldsymbol{\xi})}{\partial \xi_{1}}x_{k3} \\
        \frac{\partial N^{e}_{k}(\boldsymbol{\xi})}{\partial \xi_{2}}x_{k1} &
        \frac{\partial N^{e}_{k}(\boldsymbol{\xi})}{\partial \xi_{2}}x_{k2} &
        \frac{\partial N^{e}_{k}(\boldsymbol{\xi})}{\partial \xi_{2}}x_{k3} \\
        \frac{\partial N^{e}_{k}(\boldsymbol{\xi})}{\partial \xi_{3}}x_{k1} &
        \frac{\partial N^{e}_{k}(\boldsymbol{\xi})}{\partial \xi_{3}}x_{k2} &
        \frac{\partial N^{e}_{k}(\boldsymbol{\xi})}{\partial \xi_{3}}x_{k3}
    \end{bmatrix}
    \label{eq:fe-jac}
\end{eqnarray}
using the isoparametric transformation~\eqref{eq:fe-xe}. 
Substituting equation~\eqref{eq:fe-dv} into equation~\eqref{eq:fe-nnt-1}, the 
volume integral is expressed in the local coordinate system:
\begin{eqnarray}
	\delta\hat{\bf a}_{\mathrm{t}}({\bf x}_{0},\omega) \simeq 
	\sum\limits_{e=1}^{n_{e}}
	G \int\limits_{-1}^{+1} \! \int\limits_{-1}^{+1} \! \int\limits_{-1}^{+1} 
	\frac{\rho_{e}}{\nrm{{\bf x}-{\bf x}_{0}}^{3}}
	\left( {\bf I}_{3} - 3 {\bf e}_{\mathrm{r}} \otimes {\bf e}_{\mathrm{r}} \right)
	\hat{\bf u}^{e}(\mathbf{\boldsymbol{\xi}},\omega)	
	\det({\bf J}^{e}(\boldsymbol{\xi}))
	\, \mathrm{d}\xi_{1}
    \, \mathrm{d}\xi_{2}
    \, \mathrm{d}\xi_{3} \,,
    \label{eq:fe-nnt-2}
\end{eqnarray}
where the explicit dependence of the coordinate ${\bf x}$ on the local coordinate
$\boldsymbol{\xi}$ (equation~\eqref{eq:fe-xe}) is omitted for brevity. 
Substituting the approximation~\eqref{eq:fe-ue} of the displacement field and factoring
the element displacement vector $\underline{\hat{\bf u}}^{e}$ results in the following:
\begin{eqnarray}
	\delta\hat{\bf a}_{\mathrm{t}}({\bf x}_{0},\omega) \simeq 
	\sum\limits_{e=1}^{n_{e}}
	\left[
	G \int\limits_{-1}^{+1} \! \int\limits_{-1}^{+1} \! \int\limits_{-1}^{+1}
	\frac{\rho_{e}}{\nrm{{\bf x}-{\bf x}_{0}}^{3}}
	\left( {\bf I}_{3} - 3 {\bf e}_{\mathrm{r}} \otimes {\bf e}_{\mathrm{r}} \right)
	{\bf N}^{e}(\boldsymbol{\xi})	
	\det({\bf J}^{e}(\boldsymbol{\xi}))
	\, \mathrm{d}\xi_{1}
    \, \mathrm{d}\xi_{2}
    \, \mathrm{d}\xi_{3}
	\right]
	\underline{\hat{\bf u}}^{e} \,.
    \label{eq:fe-nnt-3}
\end{eqnarray}
The unit vector $\mathbf{e}_\mathrm{r}$ depends on the coordinate ${\bf x}$ and, hence, 
also on the local coordinate $\boldsymbol{\xi}$. 
The integrals are evaluated using Gaussian quadrature with $n_{\mathrm{G}}$ Gauss 
points with local coordinates $\boldsymbol{\xi}_{j}$ and the corresponding weights 
$w_{j}$, resulting in the following:
\begin{eqnarray}
	\delta\hat{\bf a}_{\mathrm{t}}({\bf x}_{0},\omega) & \simeq & 
    \sum\limits_{e=1}^{n_{e}}
	\left[
	G \sum\limits_{j=1}^{n_{\mathrm{G}}}
	w_{j}
	\frac{\rho_{e}}{\nrm{{\bf x}_{j}-{\bf x}_{0}}^{3}}
	\left( {\bf I}_{3} - 3 {\bf e}_{\mathrm{r}} \otimes{\bf e}_{\mathrm{r}} \right)
	{\bf N}^{e}(\boldsymbol{\xi}_{j})
	\det({\bf J}^{e}(\boldsymbol{\xi}_{j}))
	\right]
	\underline{\hat{\bf u}}^{e}
	\nonumber \\ &=&
	\sum\limits_{e=1}^{n_{e}}
	{\bf A}_{\mathrm{t}}^{e}\underline{\hat{\bf u}}^{e} \,,
    \label{eq:fe-ntt-4}
\end{eqnarray}
where ${\bf x}_{j} = {\bf x}(\boldsymbol{\xi}_{j})$ is the global coordinate of 
the Gauss point $j$. 
The matrix ${\bf A}_{\mathrm{t}}^{e}$ of dimension $3\times 3n$ yields the 
contribution of the nodal displacements $\underline{\hat{\bf u}}^{e}$ 
of element $e$ to the total Newtonian noise. 
The element matrices ${\bf A}_{\mathrm{t}}^{e}$ are assembled into a global 
matrix ${\bf A}_{\mathrm{t}}$ of dimension $3\times 3N$, with $N$ the number of 
nodes in the volume $\bar{\Omega}$, yielding:
\begin{eqnarray}
	\delta\hat{\bf a}_{\mathrm{t}}({\bf x}_{0},\omega) &=&
	{\bf A}_{\mathrm{t}} \underline{\hat{\bf u}}(\omega) \,.
    \label{eq:fe-at}
\end{eqnarray}
The displacement vector $\underline{\hat{\bf u}}(\omega)$ contains the nodal values 
of the seismic wave field in the volume~$\bar{\Omega}$.
As the matrix ${\bf A}_{\mathrm{t}}$ does not depend on the seismic wave field 
$\hat{\bf u}({\bf x},\omega)$, it only needs to be computed once for a particular 
soil layering and cavity geometry. 
The computation of the total Newtonian noise hence reduces to a matrix-vector
multiplication for a particular seismic wave field.

\subsection{Bulk contribution to the Newtonian noise}

The bulk contribution to the Newtonian noise
$\delta\hat{\bf a}_{\mathrm{b}}({\bf x}_{0},\omega)$, as defined in equation
(\ref{eq:nnbf1}), is approximated as follows with a finite element formulation:
\begin{eqnarray}
	\delta\hat{\bf a}_{\mathrm{b}}({\bf x}_{0},\omega) & \simeq &
	\sum\limits_{e=1}^{n_{e}}
	-G\int_{\bar{\Omega}^{e}} 
	\rho_{e} \boldsymbol{\chi}({\bf x},{\bf x}_{0}) 
	\nabla \cdot \hat{\bf u}^{e}({\bf x},\omega)
	\,\mathrm{d}v \,.
    \label{eq:fe-nnb-1}
\end{eqnarray}
Substituting equations~\eqref{eq:fe-ue} and~\eqref{eq:fe-dv} gives:
\begin{eqnarray}
	\delta\hat{\bf a}_{\mathrm{b}}({\bf x}_{0},\omega) & \simeq &
	\sum\limits_{e=1}^{n_{e}}
	\left[
    -G
	\int\limits_{-1}^{+1} \! \int\limits_{-1}^{+1} \! \int\limits_{-1}^{+1} 
	\rho_{e} \boldsymbol{\chi}({\bf x},{\bf x}_{0})
   {\bf B}^{e}(\boldsymbol{\xi}) 
	\det({\bf J}^{e}(\boldsymbol{\xi}))
	\, \mathrm{d}\xi_{1} 
	\, \mathrm{d}\xi_{2} 
	\, \mathrm{d}\xi_{3}
	\right]
	\underline{\hat{\bf u}}^{e} \,,
    \label{eq:fe-nnb-2}
\end{eqnarray}
where the explicit dependence of the coordinate ${\bf x}$ on the local coordinate
$\boldsymbol{\xi}$ is omitted for brevity and 
the row vector ${\bf B}^{e}(\boldsymbol{\xi})$ of dimension $1 \times 3n$ contains the
derivatives of the shape functions $N^{e}_{k}(\boldsymbol{\xi})$ with respect to 
the global coordinates ${\bf x}$:
\begin{eqnarray}
	{\bf B}^{e}(\boldsymbol{\xi}) &=&
    \left\{ \begin{array}{cccccccc}
		\frac{\partial N_{1}^{e}(\boldsymbol{\xi})}{\partial x_{1}} &
		\frac{\partial N_{1}^{e}(\boldsymbol{\xi})}{\partial x_{2}} &
		\frac{\partial N_{1}^{e}(\boldsymbol{\xi})}{\partial x_{3}} & 
		\frac{\partial N_{2}^{e}(\boldsymbol{\xi})}{\partial x_{1}} & 
		\cdots &
		\frac{\partial N_{n}^{e}(\boldsymbol{\xi})}{\partial x_{3}}
	   \end{array} \right\} \,.
\end{eqnarray}
These derivatives are computed using the inverse of the Jacobian matrix
${\bf J}^{e}(\boldsymbol{\xi})$: 
\begin{eqnarray}
    \left\{ \begin{array}{c}
	    \frac{\partial N^{e}_{k}(\boldsymbol{\xi})}{\partial x_{1}} \\
        \frac{\partial N^{e}_{k}(\boldsymbol{\xi})}{\partial x_{2}} \\
        \frac{\partial N^{e}_{k}(\boldsymbol{\xi})}{\partial x_{3}}
	  \end{array} \right\} &=&
	{\bf J}^{e}(\boldsymbol{\xi})^{-1}
    \left\{ \begin{array}{c}
	    \frac{\partial N^{e}_{k}(\boldsymbol{\xi})}{\partial \xi_{1}} \\
        \frac{\partial N^{e}_{k}(\boldsymbol{\xi})}{\partial \xi_{2}} \\
        \frac{\partial N^{e}_{k}(\boldsymbol{\xi})}{\partial \xi_{3}}
	  \end{array} \right\} \,.
\end{eqnarray}
Gaussian quadrature is applied to evaluate the volume integral in the local 
coordinate system:
\begin{eqnarray}
	\delta\hat{\bf a}_{\mathrm{b}}({\bf x}_{0},\omega) \simeq&
	\sum\limits_{e=1}^{n_{e}}
	\left[
    -G
	\sum\limits_{j=1}^{n_{\mathrm{G}}}
	w_{j} \rho_{e} \boldsymbol{\chi}({\bf x}_{j},{\bf x}_{0})
	{\bf B}^{e}(\boldsymbol{\xi}_{j})
	\det({\bf J}^{e}(\boldsymbol{\xi}_{j}))
	\right]
	\underline{\hat{\bf u}}^{e}
    \, = \,
	\sum\limits_{e=1}^{n_{e}} 
	{\bf A}_{\mathrm{b}}^{e} \underline{\hat{\bf u}}^{e} \,.
    \label{eq:fe-nnb-3}
\end{eqnarray}
Assembling the matrix ${\bf A}_{\mathrm{b}}$ of dimension $3\times 3N$ that 
accounts for the contribution of all elements yields:
\begin{eqnarray}
	\delta\hat{\bf a}_{\mathrm{b}}({\bf x}_{0},\omega) &=&
	{\bf A}_{\mathrm{b}} \underline{\hat{\bf u}}(\omega) \,.
    \label{eq:fe-ab}
\end{eqnarray}
The bulk contribution to the Newtonian noise can be computed as a matrix-vector 
multiplication for a particular seismic wave field $\underline{\hat{\bf u}}(\omega)$.

\subsection{Surface contribution to the Newtonian noise}

The surface contribution to the Newtonian noise
$\delta\hat{\bf a}_{\mathrm{s}}({\bf x}_{0},\omega)$, as defined in equation
(\ref{eq:nnsf1}), is computed by considering the surface $\bar{\Gamma}^{e}$ of each
finite element with volume $\bar{\Omega}^{e}$ individually and adding the contribution of 
all elements: 
\begin{eqnarray}
	\delta\hat{\bf a}_{\mathrm{s}}({\bf x}_{0},\omega) & \simeq &
	\sum\limits_{e=1}^{n_{e}}
    G
	\int_{\bar{\Gamma}^{e}}
	\rho_{e} \boldsymbol{\chi}({\bf x},{\bf x}_{0}) 
	\hat{\bf u}^{e}({\bf x},\omega) \cdot {\bf n}^{e}({\bf x})
	\,\mathrm{d}a  \,.
	\label{eq:fe-nns-1}
\end{eqnarray}
The contribution of a shared surface between two neighboring finite elements with 
identical density $\rho_{e}$ cancels as the unit outward normal vectors 
${\bf n}^{e}({\bf x})$ on both surfaces point in opposite directions. 
However, if the density varies from one element to another, this results in a
contribution to the Newtonian noise that is proportional to the difference in density. 
\par
The surface $\bar{\Gamma}^{e}$ of element $e$ is subdivided in $n_{s}$ 
faces (e.g. $n_{s}=4$ for tetrahedral elements and 
$n_{s}=6$ for brick elements). 
Equation~\eqref{eq:fe-nns-1} is rewritten as follows:
\begin{eqnarray}
	\delta\hat{\bf a}_{\mathrm{s}}({\bf x}_{0},\omega) & \simeq &
	\sum\limits_{e=1}^{n_{e}}
	\sum\limits_{l=1}^{n_{s}}
    G
	\int_{\bar{\Gamma}_{l}^{e}}
	\rho_{e} \boldsymbol{\chi}({\bf x},{\bf x}_{0})
	\hat{\bf u}_{l}^{e}({\bf x},\omega) \cdot {\bf n}_{l}^{e}({\bf x})
	\,\mathrm{d}a \,,
   \label{eq:fe-nns-2}
\end{eqnarray}
where $\bar{\Gamma}_{l}^{e}$ denotes the surface of the face $l$ belonging to 
element $e$, and $\hat{\bf u}_{l}^{e}({\bf x},\omega)$ and ${\bf n}_{l}^{e}({\bf x})$ 
the corresponding displacement vector and unit outward normal vector. 
The displacements $\hat{\bf u}^{e}_{l}({\bf x},\omega)$ are approximated using 
shape functions ${\bf N}^{e}(\boldsymbol{\eta})$, where the isoparametric 
transformation for surface integrals is performed from the global coordinate 
${\bf x}$ to the local coordinate 
$\boldsymbol{\eta}=\{ \eta_{1},\eta_{2} \}^{\mathrm{T}}$:
\begin{eqnarray}
	\hat{\bf u}^{e}_{l}(\boldsymbol{\eta},\omega) & \simeq &
	\sum\limits_{k=1}^{n}
	{\bf N}^{e}_{k}(\boldsymbol{\eta}) \underline{\hat{\bf u}}^{e}_{lk}
	\, = \,
	{\bf N}^{e}(\boldsymbol{\eta}) \underline{\hat{\bf u}}^{e}_{l}
	\label{eq:fe-ue-surf}
\end{eqnarray}
and $n$ is the number of nodes on the surface $\bar{\Gamma}^{e}_{l}$.
The Jacobian ${\bf J}^{e}(\boldsymbol{\eta})$ for this isoparametric 
transformation is the following $2\times 3$ matrix:
\begin{eqnarray}
    {\bf J}^{e}(\boldsymbol{\eta}) &=&
    \begin{bmatrix}
        \frac{\partial {\bf x}}{\partial \eta_{1}} \\
        \frac{\partial {\bf x}}{\partial \eta_{2}}
    \end{bmatrix}
	\, = \,
	\sum\limits_{k=1}^{n}
    \begin{bmatrix}
        \frac{\partial N^{e}_{k}(\boldsymbol{\eta})}{\partial \eta_{1}} x_{k1} &
        \frac{\partial N^{e}_{k}(\boldsymbol{\eta})}{\partial \eta_{1}} x_{k2} &
        \frac{\partial N^{e}_{k}(\boldsymbol{\eta})}{\partial \eta_{1}} x_{k3} \\
        \frac{\partial N^{e}_{k}(\boldsymbol{\eta})}{\partial \eta_{2}} x_{k1} &
        \frac{\partial N^{e}_{k}(\boldsymbol{\eta})}{\partial \eta_{2}} x_{k2} &
        \frac{\partial N^{e}_{k}(\boldsymbol{\eta})}{\partial \eta_{2}} x_{k3}
    \end{bmatrix}
	\, = \,
    \begin{bmatrix}
        {\bf j}_{1}(\boldsymbol{\eta}) \\
        {\bf j}_{2}(\boldsymbol{\eta})
    \end{bmatrix} \,.
\end{eqnarray}
The surface integral over $\bar{\Gamma}^{e}_{l}$ in equation~\eqref{eq:fe-nns-2} 
is transformed to the local coordinate $\boldsymbol{\eta}$ using the following expression for the infinitesimal area $\mathrm{d}a$:
\begin{eqnarray}
    \mathrm{d}a &=&
    \nrm{{\bf j}_{1}(\boldsymbol{\eta}) \times {\bf j}_{2}(\boldsymbol{\eta})}
    \mathrm{d}\eta_{1} \, \mathrm{d}\eta_{2} \,.
    \label{eq:fe-da}
\end{eqnarray}
Substituting expressions~\eqref{eq:fe-ue-surf} and~\eqref{eq:fe-da} into 
equation~\eqref{eq:fe-nns-2} yields:
\begin{eqnarray}
	\delta\hat{\bf a}_{\mathrm{s}}({\bf x}_{0},\omega) & \simeq &
    \sum\limits_{e=1}^{n_{e}} 
    \sum\limits_{l=1}^{n_{s}}
    \left[
    G
    \int\limits_{-1}^{+1} \! \int\limits_{-1}^{+1}
	\rho_{e} \boldsymbol{\chi}({\bf x},{\bf x}_{0})
	{\bf n}^{e}_{l}({\bf x}) {\bf N}^{e}(\boldsymbol{\eta})	
	\nrm{{\bf j}_{1}(\boldsymbol{\eta}) \times {\bf j}_{2}(\boldsymbol{\eta})}
	\, \mathrm{d}\eta_{1}
	\, \mathrm{d}\eta_{2}
	\right]
	\underline{\hat{\bf u}}^{e}_{l} \,,
   \label{eq:fe-nns-3}
\end{eqnarray}
where the explicit dependence of ${\bf x}$ on $\boldsymbol{\eta}$ is omitted 
for brevity. 
Gaussian quadrature is applied to evaluate the surface integral in the local 
coordinate system using $n_{\mathrm{G}}$ Gauss points for each face $l$:
\begin{eqnarray}
	\delta\hat{\bf a}_{\mathrm{s}}({\bf x}_{0},\omega) & \simeq &
    \sum\limits_{e=1}^{n_{e}}
    \sum\limits_{l=1}^{n_{s}}
    \left[
    G
    \sum\limits_{j=1}^{n_{\mathrm{G}}}
    w_{j}
	\rho_{e}\boldsymbol{\chi}({\bf x}_{j},{\bf x}_{0})
	{\bf n}^{e}_{l}({\bf x}_{j}) {\bf N}^{e}(\boldsymbol{\eta}_{j})	
	\nrm{{\bf j}_{1}(\boldsymbol{\eta}_{j}) \times {\bf j}_{2}(\boldsymbol{\eta}_{j})}
	\right]
	\underline{\hat{\bf u}}^{e}_{l}
	\nonumber \\ &=&
    \sum\limits_{e=1}^{n_{e}}
    \sum\limits_{l=1}^{n_{s}}
    {\bf A}^{e}_{\mathrm{s}l} \underline{\hat{\bf u}}^{e}_{l}
	\nonumber \\ &=&
    \sum\limits_{e=1}^{n_{e}} 
    {\bf A}^{e}_{\mathrm{s}} \underline{\hat{\bf u}}^{e} \,,
   \label{eq:fe-nns-4}
\end{eqnarray}
where ${\bf x}_{j} = {\bf x}(\boldsymbol{\eta}_{j})$ is the global coordinate of the Gauss 
point $j$. 
Assembling the global matrix ${\bf A}_{\mathrm{s}}$ of dimension $3\times 3N$  
to account for the contribution of all elements yields:
\begin{eqnarray}
	\delta\hat{\bf a}_{\mathrm{s}}({\bf x}_{0},\omega) &=&
	{\bf A}_{\mathrm{s}} \underline{\hat{\bf u}}(\omega) \,.
    \label{eq:fe-as}
\end{eqnarray}
The contribution of a particular part of the surface $\bar{\Gamma}$ can be 
evaluated by selecting the corresponding degrees of freedom of the nodal 
displacement vector $\underline{\hat{\bf u}}(\omega)$ and the matrix 
${\bf A}_{\mathrm{s}}$.

\subsection{Adaptive integration}

\begin{figure}[!htb]
	\centering
	(a) \includegraphics[width=0.20\textwidth]{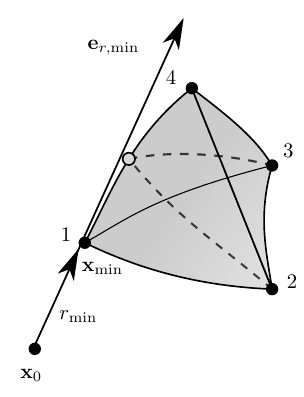} \quad
    (b) \includegraphics[width=0.20\textwidth]{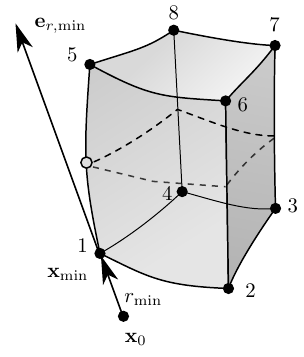} 
	\caption{Subdivision of a (a) tetrahedral and (b) hexahedral finite element into two sub-elements based on a bisection of the longest projected length.}
	\label{fig:adaptive-int}
\end{figure}

The function
$\boldsymbol{\chi}({\bf x},{\bf x}_{0})$ in
equations~\eqref{eq:nnbf1} and~\eqref{eq:nnsf1} 
and its gradient in equation~(\ref{eq:nntf1}) 
exhibit a strong spatial variation in the vicinity of the mirror position ${\bf x}_{0}$ due to the singular factor $1/{\nrm{{\bf x} - {\bf x}_{0}}^{3}}$. Therefore, the element matrices ${\bf A}_{\mathrm{t}}^{e}$, ${\bf A}_{\mathrm{b}}^{e}$, and ${\bf A}_{\mathrm{s}}^{e}$ are evaluated using adaptive integration. Following approaches commonly used in boundary element methods for weakly singular and near-singular integrals~\cite{zhang15}, adaptive integration is employed to accurately capture the strongly varying Newtonian noise kernels in the finite element surface and volume contributions.
This proceeds by subdividing the element into sub-elements (Figure \ref{fig:adaptive-int}).
\par
For a given element, the corner node closest to the mirror position ${\bf x}_{0}$ is first identified. Denoting its position by ${\bf x}_{\min}$,
the minimum distance and corresponding radial unit vector are defined as
\begin{eqnarray}
   r_{\min} &=& \min_i \|{\bf x}_i-{\bf x}_0\| \,,
\end{eqnarray}
and
\begin{eqnarray}
   {\bf e}_{\mathrm{r},\min} &=& 
   \frac{{\bf x}_{\min}-{\bf x}_0}
   {\|{\bf x}_{\min}-{\bf x}_0\|} \,.
\end{eqnarray}
For each edge $(i,j)$ of the current integration domain, the edge length
projected onto the radial direction is computed as
\begin{eqnarray}
   l_{ij} &=& 
   \left|({\bf x}_{j}-{\bf x}_{i})\cdot{\bf e}_{\mathrm{r},\min}\right| \,.
\end{eqnarray}
The integration domain satisfies the refinement criterion if
\begin{eqnarray}
   \max(l_{ij}) &\leq& \alpha r_{\min} \,,
    \label{eq:refinementcrit}
\end{eqnarray}
where $\alpha$ is a refinement parameter. If the criterion is not satisfied, triangular and tetrahedral domains are refined by bisecting the edge with the largest projected length, whereas quadrilateral and hexahedral domains are subdivided along the local coordinate direction associated with the largest projected length. This procedure is applied recursively until all sub-elements satisfy the refinement criterion \eqref{eq:refinementcrit}, after which standard Gaussian quadrature is applied to all sub-elements.

\subsection{Implementation}
The finite element computation of the Newtonian noise is implemented in the ANNA Newtonian Noise Analysis toolbox within the MATLAB programming and numeric computing platform, making use of subroutines embedded in Stabil, an  educational MATLAB toolbox for static and dynamic structural analysis 
\cite{ij-caee-fran-21a}. ANNA is compatible with the open-source GNU Octave Scientific Programming Language \cite{quar06a}, while a Python version \cite{olip07a} is also available, making use of the high‑performance NumPy package 
for array operations and numerical kernels for arrays \cite{harri20a}.
\par
The Newtonian noise is computed for a seismic wave field that
is defined on a 3D finite element mesh. 
The toolbox in its present form supports linear (4-node) and quadratic (10-node)
tetrahedral finite elements, as well as linear (8-node) and quadratic (20-node)
brick finite elements.
\par
At the core of ANNA is a dedicated subroutine {\tt asmnn.m} that
takes the 3D finite element mesh and material properties (density) as input and
computes the finite element matrices 
${\bf A}_{\mathrm{t}}$, ${\bf A}_{\mathrm{b}}$ and ${\bf A}_{\mathrm{s}}$ of 
dimension $3 \times 3N$, with $N$ the number of nodes in the finite element mesh.

\subsection{Work flow}

The computation of Newtonian noise hence reduces to the following four-step procedure:
\begin{enumerate}
\item Create a 3D finite element mesh with an extension tailored
to represent the seismic wave field in the frequency range of interest.
The adopted mesh criteria are discussed in detail in section~\ref{sec:verification}. Gmsh~\cite{geuz09a} is used to generate the resulting 3D finite element meshes with $N$ nodes and $3N$ degrees of freedom.
\item Compute a 3D seismic wave field, stored as a column vector 
$\underline{\hat{\bf u}}(\omega)$ of dimension $3N \times 1$, where each entry 
corresponds to a degree of freedom of the $N$ nodes of a finite element mesh. 
As mentioned in the introduction, a wide variety of methods and tools with different 
degrees of complexity can be used to compute the seismic wave field.
\item Use the subroutine {\tt asmnn.m} in ANNA to compute the finite element 
matrices ${\bf A}_{\mathrm{t}}$, ${\bf A}_{\mathrm{b}}$ and ${\bf A}_{\mathrm{s}}$
of dimension $3 \times 3N$, based on the 3D finite element mesh and material
properties (density).
\item Compute the total Newtonian noise 
$\delta\hat{\bf a}_{\mathrm{t}}({\bf x}_{0},\omega)$ 
according to equation (\ref{eq:fe-at}) as the product of 
the finite element matrix ${\bf A}_{\mathrm{t}}$ and 
the displacement vector $\underline{\hat{\bf u}}(\omega)$, 
resulting in a column vector of dimension $3 \times 1$.
Similarly, the bulk and surface contributions 
$\delta\hat{\bf a}_{\mathrm{b}}({\bf x}_{0},\omega)$ and
$\delta\hat{\bf a}_{\mathrm{s}}({\bf x}_{0},\omega)$ are computed 
according to equations (\ref{eq:fe-ab}) and (\ref{eq:fe-as}) as the product of 
the finite element matrices ${\bf A}_{\mathrm{b}}$ and ${\bf A}_{\mathrm{s}}$ and 
the displacement vector $\underline{\hat{\bf u}}(\omega)$.
\end{enumerate}

\section{Verification}
\label{sec:verification}

The resulting numerical formulation is verified using available analytical solutions for Newtonian noise~\cite{harm19a}.
Three configurations are considered. 
First, plane harmonic P- and S-waves propagating in a homogeneous linear elastic
full space with a cavity with radius $r_{0}$ are examined.
Next, Rayleigh waves propagating along the surface of a homogeneous linear
elastic halfspace are considered, and the Newtonian noise on a test mass at 
a distance $h$ above the free surface is computed.

\subsection{Plane harmonic P-wave in a full space}
\label{subsec:verif-full-space-p}

First, a mirror suspended in a spherical cavity with radius $r_{0}$ at the center 
${\bf x}_{0} = \{ 0,0,0 \}^{\mathrm{T}}$ of a homogeneous linear elastic full 
space $\Omega$ is considered. 
Newtonian noise is computed for a plane harmonic P-wave with a wavelength 
$\lambda_{\mathrm{p}}$ that \textendash~in the low frequency range considered \textendash~is assumed 
to be much larger than the cavity radius $r_{0}$, so that wave scattering due to 
the cavity can be ignored.
\par
The displacement vector $\hat{\bf u}({\bf x},\omega)$ due to a plane harmonic P-wave 
with unit amplitude propagating with velocity $C_{\mathrm{p}}$ in the direction 
${\bf e}_{\mathrm{k}}$ is equal to:
\begin{eqnarray}
	\hat{\bf u}({\bf x},\omega) &=&
	\exp(-\mathrm{i}k_{\mathrm{p}} {\bf e}_{\mathrm{k}} \cdot{\bf x}) {\bf e}_{\mathrm{k}} \,,
	\label{eq:up}
\end{eqnarray}
where $k_{\mathrm{p}}=\omega/C_{\mathrm{p}}=2 \pi/\lambda_{\mathrm{p}}$ is the
longitudinal wavenumber and $\omega$ is the circular frequency.
\par
\begin{figure}[!htb]
	\centering
	(a) \includegraphics[width=0.425\textwidth]{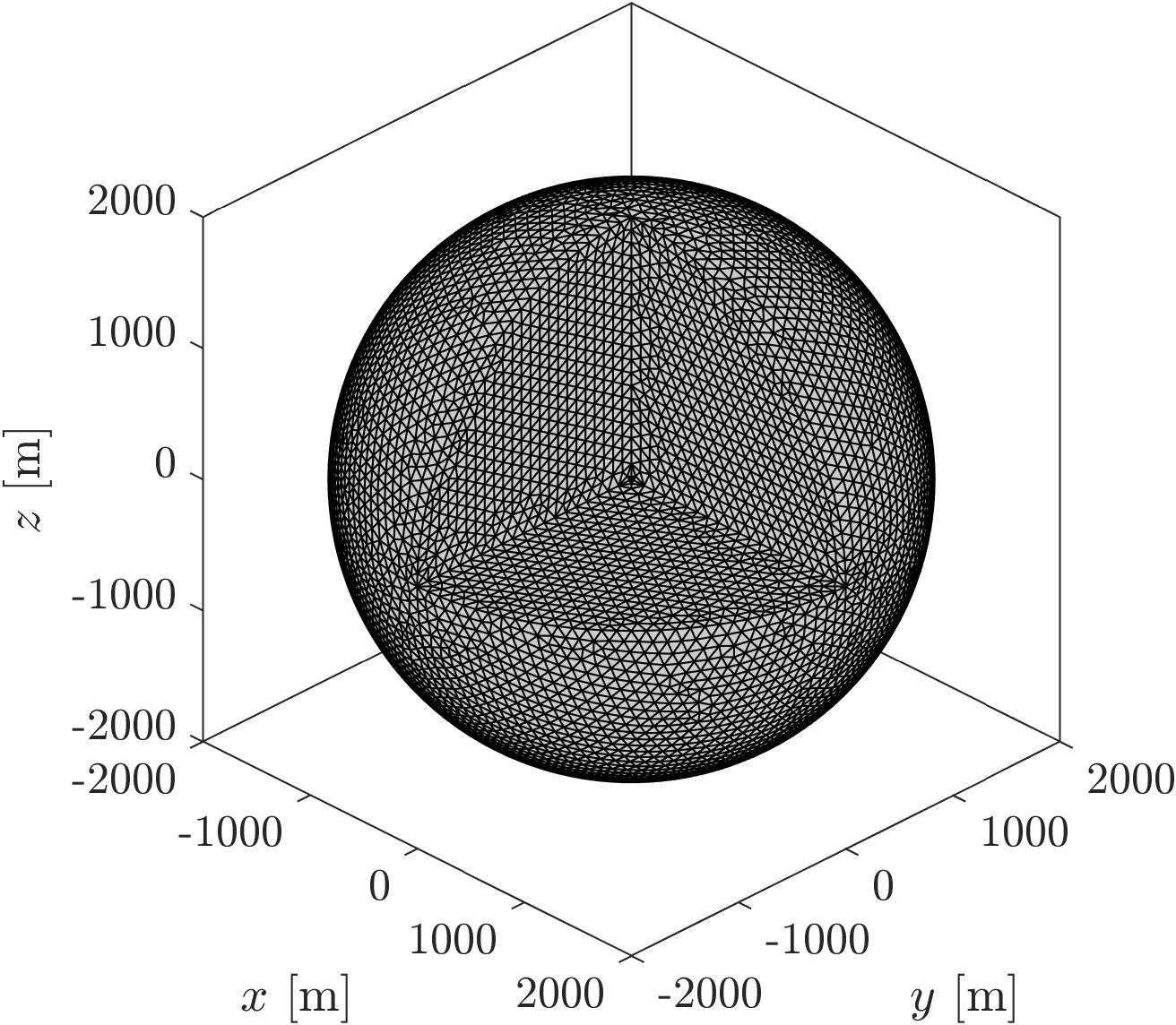} \quad
    (b) \includegraphics[width=0.425\textwidth]{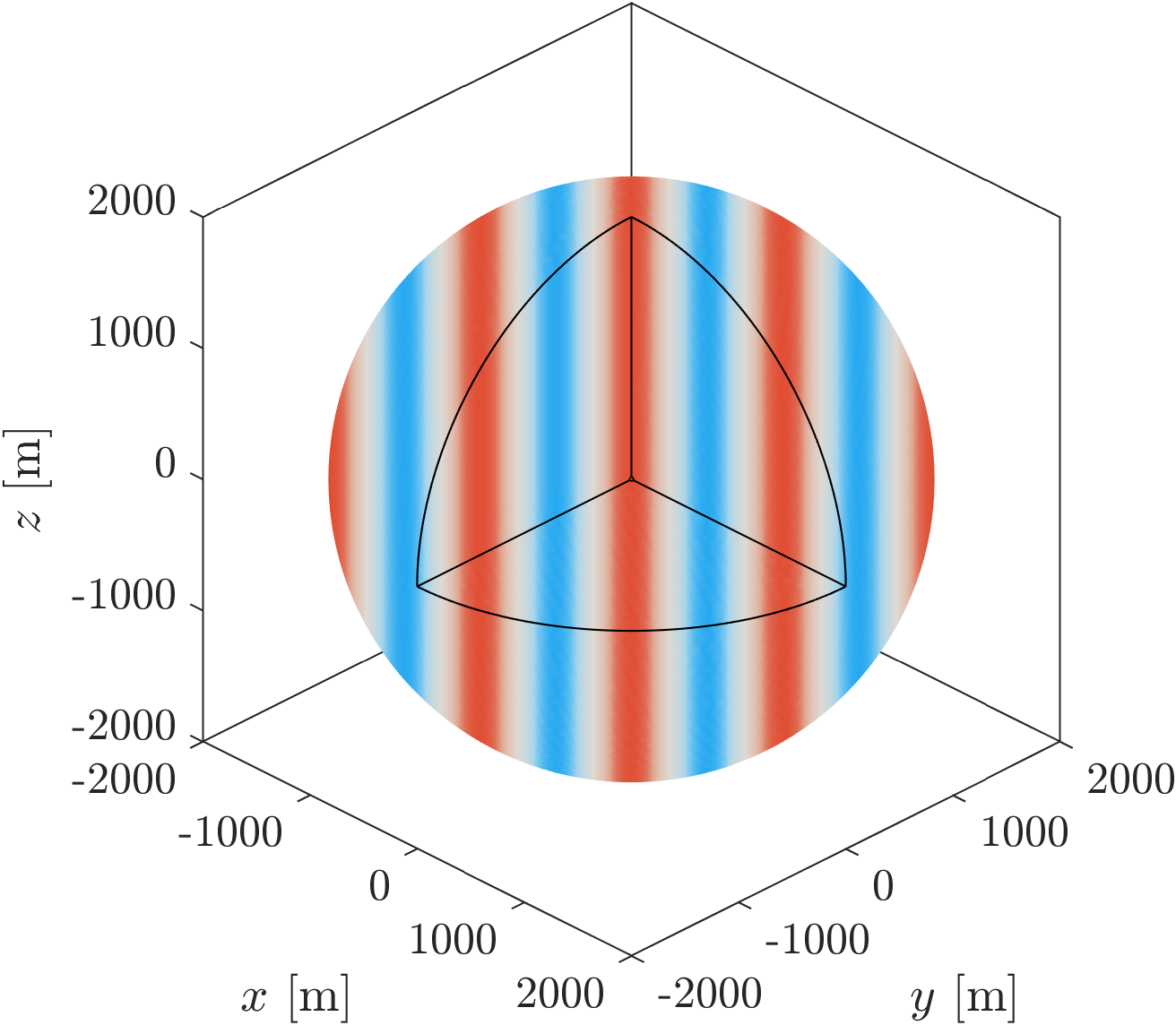} 
	\caption{(a) Finite element mesh of a sphere 
             with center at ${\bf x}_{0} = \{ 0,0,0 \}^{\mathrm{T}}$
             and radius $R=2000\,\mbox{m}$ 
             and a spherical cavity with radius $r_{0}=20\,\mbox{m}$,
             used to compute the Newtonian noise 
             due to a plane harmonic P-wave
             in a frequency range around 5\,Hz.
             (b) Displacement $\hat{u}_{x}({\bf x},\omega)$ 
             due to a plane harmonic P-wave 
             with unit amplitude and frequency of 5\,Hz,
             propagating in the direction 
             ${\bf e}_{\mathrm{k}}=\{\frac{1}{\sqrt{2}},\frac{1}{\sqrt{2}},0\}^{\mathrm{T}}$.
             }
	\label{fig:fe-mesh-full-space-p}
\end{figure}
Figure~\ref{fig:fe-mesh-full-space-p}b shows the displacement component 
$\hat{u}_{x}({\bf x},\omega)$ of a plane harmonic P-wave with unit amplitude and 
frequency of 5\,Hz, propagating in the direction 
${\bf e}_{\mathrm{k}}=\{\frac{1}{\sqrt{2}},\frac{1}{\sqrt{2}},0\}^{\mathrm{T}}$ 
in a homogeneous linear elastic full space with 
shear wave velocity $C_{\mathrm{s}}=2500 \, \mbox{m/s}$, 
dilatational wave velocity $C_{\mathrm{p}}=5000 \, \mbox{m/s}$, and 
density $\rho=2800 \,\mbox{kg/m}^{3}$;
the wave field is shown on a spherical finite element mesh that will subsequently 
be used to compute the Newtonian noise.
Since the displacement $\hat{u}_{y}({\bf x},\omega)$ is equal to 
$\hat{u}_{x}({\bf x},\omega)$, while $\hat{u}_{z}({\bf x},\omega)$ is zero, 
these components are not shown.
The wavelength $\lambda_{\mathrm{p}}=1000$~m at 5~Hz is much larger than the 
cavity radius $r_{0}=20$~m, so that wave scattering by the cavity indeed can be 
disregarded.
\par
For a plane harmonic P-wave propagating in a linear elastic full space, an 
analytical expression for the Newtonian noise can be derived if wave scattering
by the cavity is ignored.
Inserting equation~\eqref{eq:up} into equation~\eqref{eq:nntf1} and evaluating 
the volume integral in spherical coordinates between radii $r_{0}$ and $R$ 
yields the following analytical expression for the Newtonian noise 
$\delta\hat{\bf a}_{\mathrm{t}}({\bf x}_{0},\omega)$ at the mirror position 
${\bf x}_{0}$~\cite{harm19a}:
\begin{eqnarray}
	\delta\hat{\bf a}_{\mathrm{t}}({\bf x}_{0},\omega) &=&
	8 \pi \rho G
    \left(
    \frac{j_{1}(k_{\mathrm{p}} r_{0})}{k_{\mathrm{p}} r_{0}} -
    \frac{j_{1}(k_{\mathrm{p}} R)}{k_{\mathrm{p}} R}
    \right)
    \hat{\bf u}({\bf x}_{0},\omega) \,,
    \label{eq:at-p-R}
\end{eqnarray}
where $j_{1}(\cdot)$ is the spherical Bessel function of the first kind and order $1$. 
Assuming a full space, taking the limit for $R\rightarrow \infty$ yields:
\begin{eqnarray}
	\delta\hat{\bf a}_{\mathrm{t}}({\bf x}_{0},\omega) &=&
	8 \pi \rho G
    \frac{j_{1}(k_{\mathrm{p}} r_{0})}{k_{\mathrm{p}} r_{0}}     
    \hat{\bf u}({\bf x}_{0},\omega) \,.
	\label{eq:at-p-inf}
\end{eqnarray}
The analytical solution for the bulk contribution 
$\delta\hat{\bf a}_{\mathrm{b}}({\bf x}_{0},\omega)$ is
\begin{eqnarray}
    \delta\hat{\bf a}_{\mathrm{b}}({\bf x}_{0},\omega) &=&
    4 \pi \rho G
    \left(
    j_{0}(k_{\mathrm{p}} r_{0}) - 
    j_{0}(k_{\mathrm{p}} R)
    \right)
    \hat{\bf u}({\bf x}_{0},\omega) 
	\label{eq:ab-p-R}
\end{eqnarray}
and, for a full space, the corresponding limit for $R\rightarrow \infty$ is:
\begin{eqnarray}
    \delta\hat{\bf a}_{\mathrm{b}}({\bf x}_{0},\omega) &=&
    4 \pi \rho G
    j_{0}(k_{\mathrm{p}} r_{0})
    \hat{\bf u}({\bf x}_{0},\omega) \,. 
    \label{eq:ab-p-inf}
\end{eqnarray}
Similarly, the analytical solution for the surface contribution 
$\delta\hat{\bf a}_{\mathrm{s}}({\bf x}_{0},\omega)$ is:
\begin{eqnarray}
    \delta\hat{\bf a}_{\mathrm{s}}({\bf x}_{0},\omega) &=&
     8 \pi \rho G
    \left(
    \frac{j_{1}(k_{\mathrm{p}} r_{0})}{k_{\mathrm{p}} r_{0}} - 
    \frac{1}{2} j_{0}(k_{\mathrm{p}} r_{0})
    -\frac{j_{1}(k_{\mathrm{p}} R)}{k_{\mathrm{p}} R} + 
    \frac{1}{2} j_{0}(k_{\mathrm{p}} R)
    \right)
    \hat{\bf u}({\bf x}_{0},\omega) \,,
    \label{eq:as-p-R}
\end{eqnarray}
while its full space limit is given by:
\begin{eqnarray}
    \delta\hat{\bf a}_{\mathrm{s}}({\bf x}_{0},\omega) &=&
    8 \pi \rho G
    \left(
    \frac{j_{1}(k_{\mathrm{p}} r_{0})}{k_{\mathrm{p}} r_{0}} - 
    \frac{1}{2} j_{0}(k_{\mathrm{p}} r_{0})
    \right)
    \hat{\bf u}({\bf x}_{0},\omega) \,.
    \label{eq:as-p-inf}
\end{eqnarray}
\par
The Newtonian noise is subsequently evaluated 
using the finite element formulation presented in section \ref{sec:fe}. 
A finite element mesh is constructed on a spherical domain with center at ${\bf x}_{0}$ 
and radius $R$ for calculations in the frequency range $f \in [f_{\mathrm{min}},f_{\mathrm{max}}]$, 
where $f$ is equal to $\omega/(2\pi)$, taking into consideration the following criteria:
\begin{enumerate}
\item The radius $R$ of the spherical finite element domain must be 
sufficiently large in order to well incorporate the far field contributions to the 
Newtonian noise for all frequencies $f \in [f_{\mathrm{min}},f_{\mathrm{max}}]$.
\par
The truncation error $\Delta_{\mathrm{t}}(R,\omega)$ on the total Newtonian noise 
$\delta\hat{\bf a}_{\mathrm{t}}({\bf x}_{0},\omega)$ for a spherical domain with
radius $R$ is obtained by subtracting equation~\eqref{eq:at-p-R} from
equation~\eqref{eq:at-p-inf}:
\begin{eqnarray}
   \Delta_{\mathrm{t}}(R,\omega) &=& 
   +
   8\pi \rho G
   \frac{j_{1}(k_{\mathrm{p}} R)}{k_{\mathrm{p}}R}
   \hat{\bf u}({\bf x}_{0},\omega) \,.
\label{eq:at-p-delta}
\end{eqnarray}
This truncation error is governed by the factor
$\left| j_{1}(k_{\mathrm{p}}R) / (k_{\mathrm{p}}R ) \right|$, which is limited to a tolerance
$\varepsilon$ at the lowest frequency $f_{\mathrm{min}}$:
\begin{eqnarray}
   \left| 
   \frac{j_{1}(k_{\mathrm{p}} R)}{k_{\mathrm{p}}R}
   \right| \le \varepsilon
   \quad \text{for} \quad 
   f = f_{\mathrm{min}} \,.
   \label{eq:at-p-tolerance}
\end{eqnarray}
For a tolerance $\varepsilon=0.01$, a radius 
$R\gtrsim 1.7\lambda_{\mathrm{p}}^{\mathrm{max}}$ is required.
\par
The truncation error $\Delta_{\mathrm{b}}(R,\omega)$ on the bulk contribution is equal to:
\begin{eqnarray}
   \Delta_{\mathrm{b}}(R,\omega) &=&
   -4\pi\rho G  
   j_{0}(k_{\mathrm{p}}R) \hat{\mathbf u}(\mathbf x_{0},\omega) 
\label{eq:ab-p-delta}
\end{eqnarray}
and a tolerance $\varepsilon$ can again be imposed on the factor depending on $k_{\mathrm{p}}R$:
\begin{eqnarray}
   \left| 
   j_{0}(k_{\mathrm{p}}R)
   \right| \le \varepsilon
   \quad \text{for} \quad 
   f = f_{\mathrm{min}} \,.
   \label{eq:ab-p-tolerance}
\end{eqnarray}
The truncation error $\Delta_{\mathrm{s}}(R,\omega)$ on the surface contribution is equal to:
\begin{eqnarray}
   \Delta_{\mathrm{s}}(R,\omega) &=&
   + 4\pi\rho G 
   \left( 
   j_{0}(k_{\mathrm{p}}R) - 
   2 \frac{j_{1}(k_{\mathrm{p}}R)}{k_{\mathrm{p}}R}
   \right)
   \hat{\mathbf u}(\mathbf x_{0},\omega)
\label{eq:as-p-delta}
\end{eqnarray}
and a tolerance $\varepsilon$ can be imposed on the factor depending on $k_{\mathrm{p}}R$:
\begin{eqnarray}
   \left| 
   j_{0}(k_{\mathrm{p}}R) - 
   2 \frac{j_{1}(k_{\mathrm{p}}R)}{k_{\mathrm{p}}R}
   \right| \le \varepsilon
   \quad \text{for} \quad 
   f= f_{\mathrm{min}} \,.
   \label{eq:as-p-tolerance}
\end{eqnarray}
As the truncation errors $\Delta_{\mathrm{b}}(R,\omega)$ and $\Delta_{\mathrm{s}}(R,\omega)$
decay more slowly with $R$ than the truncation error $\Delta_{\mathrm{t}}(R,\omega)$,
enforcing a similar tolerance $\varepsilon=0.01$ in equations~\eqref{eq:ab-p-tolerance}
and~\eqref{eq:as-p-tolerance} would require a substantially larger domain with radius
$R\gtrsim 15.9\lambda_{\mathrm{p}}^{\mathrm{max}}$ for the bulk contribution and 
$R\gtrsim 16.2\lambda_{\mathrm{p}}^{\mathrm{max}}$ for the surface contribution. 
\item The element size $l_{\mathrm{e}}$ should be sufficiently small 
to well resolve the propagating waves. It is therefore limited to 
$l_{\mathrm{e}}=\lambda^{\mathrm{min}}/n$, with $\lambda^{\mathrm{min}}$ 
the smallest wavelength and $n$ the number of finite elements per wavelength.
\end{enumerate}
\par
Figure~\ref{fig:fe-mesh-full-space-p}a shows the finite element mesh that is used for 
the numerical computation of the Newtonian noise due to a plane harmonic P-wave or S-wave.
This finite element mesh is generated with Gmsh~\cite{geuz09a} 
on a sphere with center at ${\bf x}_{0} = \{ 0,0,0 \}^{\mathrm{T}}$ and
radius $R=2000\,\mbox{m}$ and a cavity with radius $r_{0}=20\,\mbox{m}$,
using 10–node quadratic tetrahedral finite elements. 
The frequency range of interest is defined by $f_{\mathrm{min}}=f_{\mathrm{max}}=5\,$Hz.
In order to avoid an excessive computational cost, the domain size is selected as
$R=2\lambda_{\mathrm{p}}^{\mathrm{max}}$.
According to equation~\eqref{eq:at-p-tolerance}, this choice provides a sufficiently 
low tolerance of $1/(16\pi^{2})$ on the total Newtonian noise. 
According to equations~\eqref{eq:ab-p-tolerance} and~\eqref{eq:as-p-tolerance},
however, the truncation errors on the bulk and surface contributions are expected
to be larger. 
As the same finite element mesh is used for S- and P-wave propagation, the
finite element size $l_{\mathrm{e}}$  is defined as 
$\lambda_{\mathrm{s}}^{\mathrm{min}}/n$ with the minimum shear wavelength 
$\lambda_{\mathrm{s}}^{\mathrm{min}}=C_\mathrm{s}/f_\mathrm{max}$  
and $n=5$.
The adaptive integration parameter is selected as $\alpha=0.1$.
This value was found to provide a suitable compromise between computational efficiency and accuracy for the examples considered.
\par
\begin{figure}[!htb]
	\centering
	(a) \includegraphics[width=0.425\textwidth]{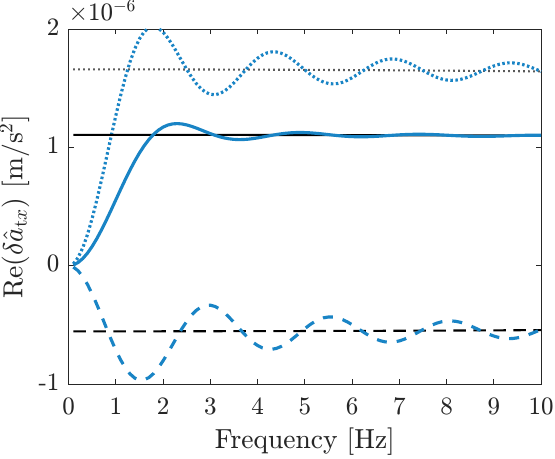} \quad
	(b) \includegraphics[width=0.425\textwidth]{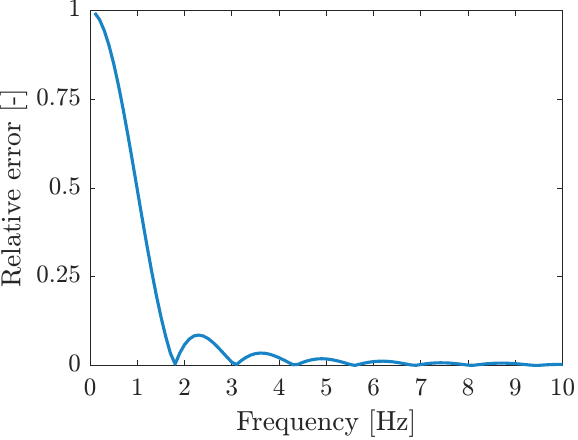}
	\caption{(a) Real part of the total Newtonian noise 
             $\delta\hat{a}_{\mathrm{t}x}({\bf x}_{0},\omega)$ 
             computed analytically ($\textcolor{black}{\full}$) and 
             with the numerical model ($\textcolor{blue2}{\full}$), 
             due to a plane harmonic P-wave in a linear elastic full space
             with a cavity with radius $r_{0}=20\,\mbox{m}$;
             the P-wave has unit amplitude and propagates in the direction 
             ${\bf e}_{\mathrm{k}}=\{\frac{1}{\sqrt{2}},\frac{1}{\sqrt{2}},0\}^{\mathrm{T}}$.
             The bulk contribution 
             $\delta\hat{a}_{\mathrm{b}x}({\bf x}_{0},\omega)$ 
             ($\textcolor{black}{\dotted}$ and $\textcolor{blue2}{\dotted}$) and 
             the surface contribution 
             $\delta\hat{a}_{\mathrm{s}x}({\bf x}_{0},\omega)$ 
             ($\textcolor{black}{\dashed}$ and $\textcolor{blue2}{\dashed}$)
             are also shown.           
             (b) Relative error $\varepsilon_{\mathrm{t}x}(\omega)$ 
             on the total Newtonian noise.
             }
	\label{fig:verif-full-space-p}
\end{figure}

Figure~\ref{fig:verif-full-space-p}a compares the real part of the total 
Newtonian noise 
$\delta\hat{a}_{\mathrm{t}x}^{\mathrm{num}}({\bf x}_{0},\omega)$ 
computed with the finite element approximation with the real part of the analytical solution 
$\delta\hat{a}_{\mathrm{t}x}^{\mathrm{ref}}({\bf x}_{0},\omega)$ 
in the frequency range between 0.1\,Hz and 10\,Hz.
The displacement $\hat{\bf u}({\bf x}_{0},\omega)$ at the mirror in 
equation~(\ref{eq:at-p-inf}) is purely real,
so that $\delta\hat{a}_{\mathrm{t}x}({\bf x}_{0},\omega)$ is also real.
The imaginary part vanishes and therefore is not shown. 
The difference between the numerical and reference solution is expressed by 
the relative error: 
\begin{eqnarray}
   \varepsilon_{\mathrm{t}x}(\omega) &=&
    \frac{\left|\delta\hat{a}_{\mathrm{t}x}^{\mathrm{num}}({\bf x}_{0},\omega) -
               \delta\hat{a}_{\mathrm{t}x}^{\mathrm{ref}}({\bf x}_{0},\omega)\right|}
         {\left|\delta\hat{a}_{\mathrm{t}x}^{\mathrm{ref}}({\bf x}_{0},\omega)\right|}\,,
   \label{eq:at-p-error}
\end{eqnarray}
where $|\cdot|$ denotes the magnitude of a complex number.

In the low frequency range considered, the total Newtonian noise 
$\delta\hat{a}_{\mathrm{t}x}^{\mathrm{ref}}({\bf x}_{0},\omega)$ is nearly constant.
The numerical prediction 
$\delta\hat{a}_{\mathrm{t}x}^{\mathrm{num}}({\bf x}_{0},\omega)$ 
agrees very well with the analytical result at frequencies larger than 5\,Hz. 
At these frequencies, the extension of the finite element mesh is sufficiently large, while the finite element size remains small enough to capture the wavelength.
This is reflected by a low relative error $\varepsilon_{\mathrm{t}x}(\omega)$ in 
figure~\ref{fig:verif-full-space-p}b.
For lower frequencies, however, the numerical solution deviates from the reference 
solution due to the fixed domain size $R=2\lambda_{\mathrm{p}}^{\mathrm{max}}$. 
The condition in equation~\eqref{eq:at-p-tolerance} is no longer satisfied, and a 
larger domain size $R$ would be required to maintain the same level of accuracy.
The total Newtonian noise computed numerically from equation~\eqref{eq:fe-at} is, 
as expected, equal to the sum of the bulk and surface contributions obtained from 
equations~\eqref{eq:fe-ab} and~\eqref{eq:fe-as}.
The predicted bulk and surface contributions do not converge independently to their analytical values. This is expected, since the tolerance criterion in equation~\eqref{eq:at-p-tolerance} was applied only to the total Newtonian noise and not to the separate contributions.

\begin{figure}[!htb]
	\centering
	\includegraphics[width=0.425\textwidth]{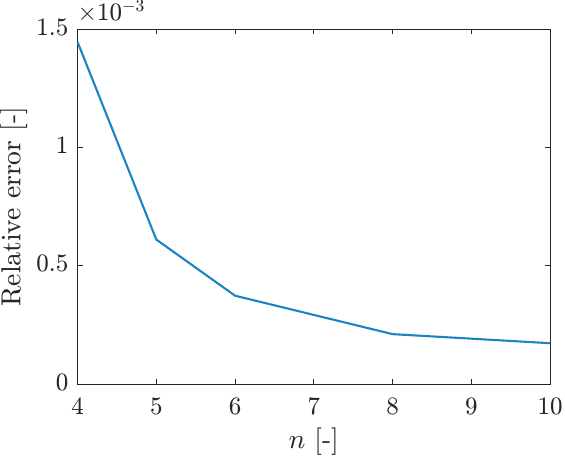} 
	\caption{Relative error $\varepsilon_{\mathrm{t}x}$ 
    with respect to the analytical solution for a finite spherical domain with radius $R=2000\,\mbox{m}$, as a function of the number of finite elements per minimum shear wavelength $n$ at $f=5\,\mbox{Hz}$.
             }
	\label{fig:verif-full-space-p-convergence}
\end{figure}

Figure~\ref{fig:verif-full-space-p-convergence} shows the relative error of the total Newtonian noise at $f=5\,\mbox{Hz}$ for different values of the number of finite
elements per minimum shear wavelength $n=4,\,5,\,6,\,8,\,10$.
The relative error decreases with increasing number of finite elements per wavelength and becomes nearly constant for $n\geq6$.
For the value $n=5$ adopted throughout this work, the relative error is below $10^{-3}$.
Further mesh refinement produces only marginal changes in the numerical prediction,
indicating that the adopted discretization provides sufficient accuracy for the
examples considered.

\subsection{Plane harmonic S-wave in a full space}
\label{subsec:verif-full-space-s}

Newtonian noise is subsequently computed for a plane harmonic S-wave propagating
in a homogeneous elastic full space with a wavelength $\lambda_{\mathrm{s}}$ that 
\textendash~in the low frequency range considered \textendash~is assumed to be 
much larger than the cavity radius $r_{0}$, so that wave scattering due to the 
cavity can again be ignored.
\par
The displacement vector $\hat{\bf u}({\bf x},\omega)$ due to a plane harmonic S-wave 
with unit amplitude propagating with velocity $C_{\mathrm{s}}$ in the direction 
${\bf e}_{\mathrm{k}}$ and with polarization vector ${\bf e}_{\mathrm{s}}$ is equal to:
\begin{eqnarray}
	\hat{\bf u}({\bf x},\omega) &=&
	\exp(-\mathrm{i}k_{\mathrm{s}} {\bf e}_{\mathrm{k}} \cdot{\bf x}) {\bf e}_{\mathrm{s}}\,,
	\label{eq:us}
\end{eqnarray}
where $k_{\mathrm{s}}=\omega/C_{\mathrm{s}}=2 \pi/\lambda_{\mathrm{s}}$ is the
shear wavenumber and $\lambda_{\mathrm{s}}$ is the corresponding wavelength.
The polarization vector ${\bf e}_{\mathrm{s}}$ is orthogonal to the propagation 
direction ${\bf e}_{\mathrm{k}}$.
\par
\begin{figure}[!htb]
	\centering
	(a) \includegraphics[width=0.425\textwidth]{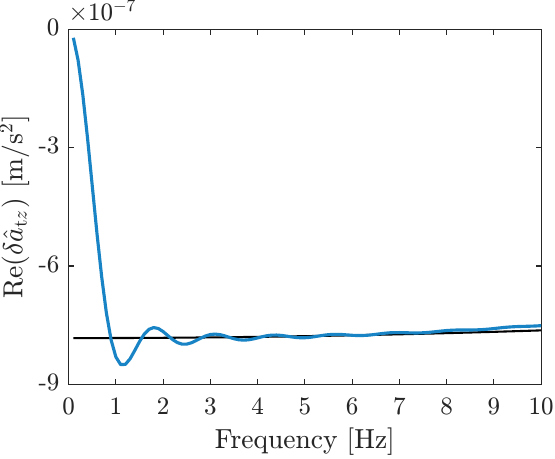} \quad
	(b) \includegraphics[width=0.425\textwidth]{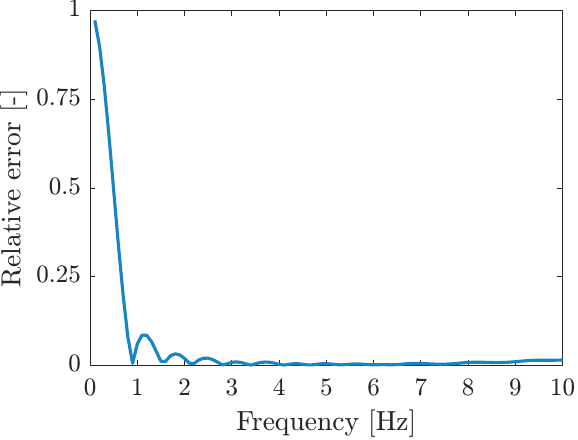}
	\caption{(a) Real part of the total Newtonian noise 
             $\delta\hat{a}_{\mathrm{t}z}({\bf x}_{0})$ 
             computed analytically ($\textcolor{black}{\full}$) and 
             with the numerical model ($\textcolor{blue2}{\full}$), 
             due to a plane harmonic S-wave in a linear elastic full space
             with a cavity with radius $r_{0}=20\,\mbox{m}$;
             the S-wave has unit amplitude and propagates in the direction 
             ${\bf e}_{\mathrm{k}}=\{1,0,0\}^{\mathrm{T}}$
             with polarization vector
             ${\bf e}_{\mathrm{s}}=\{0,0,1\}^{\mathrm{T}}$.
             (b) Relative error $\varepsilon_{\mathrm{t}z}(\omega)$ 
             on the total Newtonian noise.
             }
	\label{fig:verif-full-space-s}
\end{figure}

Figure~\ref{fig:verif-full-space-s}b shows the displacement component 
$\hat{u}_{z}({\bf x},\omega)$ of a plane harmonic S-wave with unit amplitude and frequency 
of 5\,Hz, propagating in the direction ${\bf e}_{\mathrm{k}}=\{1,0,0\}^{\mathrm{T}}$
with polarization vector ${\bf e}_{\mathrm{s}}=\{0,0,1\}^{\mathrm{T}}$
in a homogeneous linear elastic full space with properties as introduced before;
the wave field is shown on a finite element mesh that will subsequently 
be used to compute the Newtonian noise.
The wavelength $\lambda_{\mathrm{s}}=500$~m at 5~Hz is much larger than the 
cavity radius $r_{0}=20$~m, so that wave scattering by the cavity again can be 
disregarded.
\begin{figure}[!htb]
	\centering
    	(a) \includegraphics[width=0.425\textwidth]{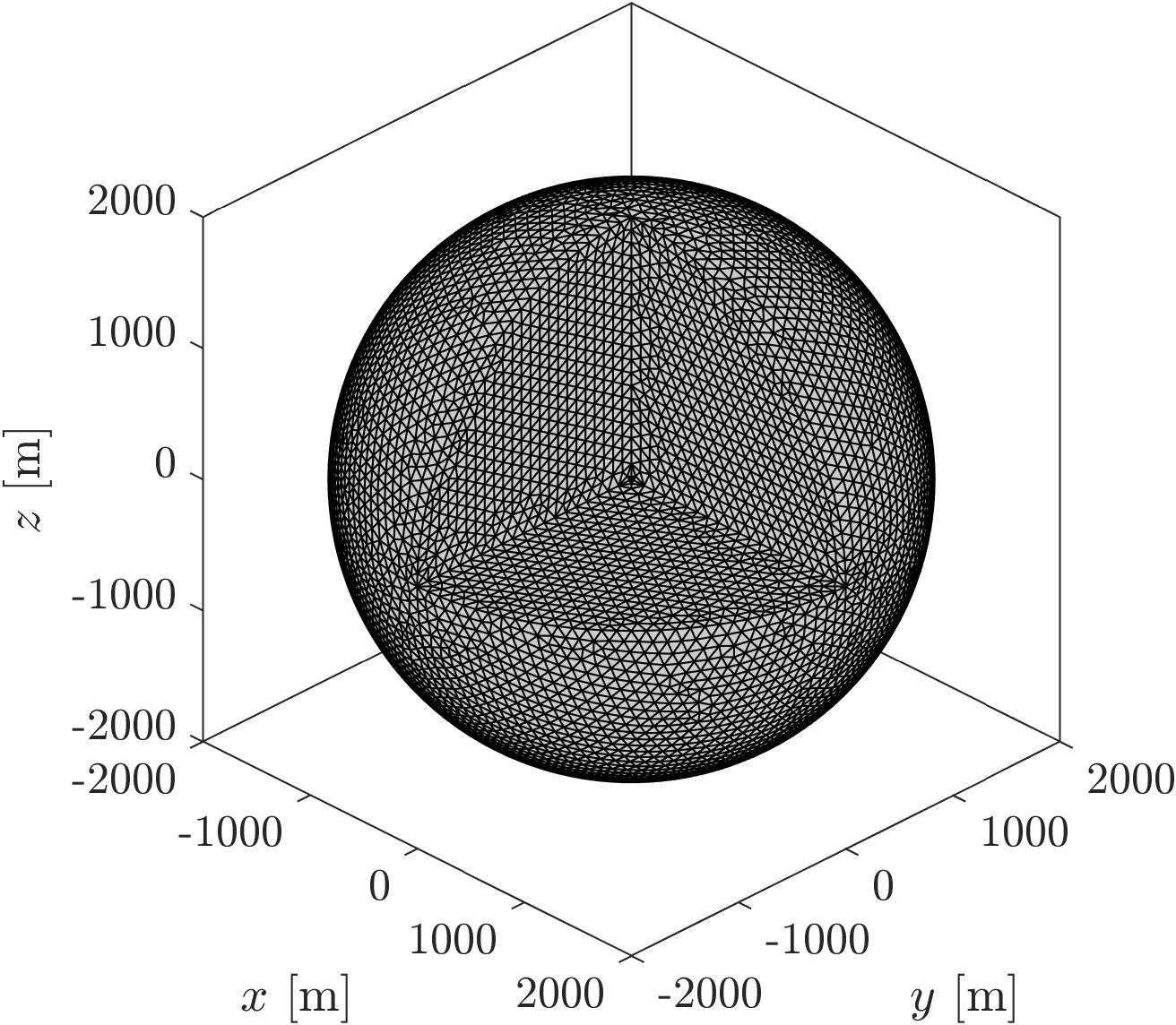} \quad
    (b) \includegraphics[width=0.425\textwidth]{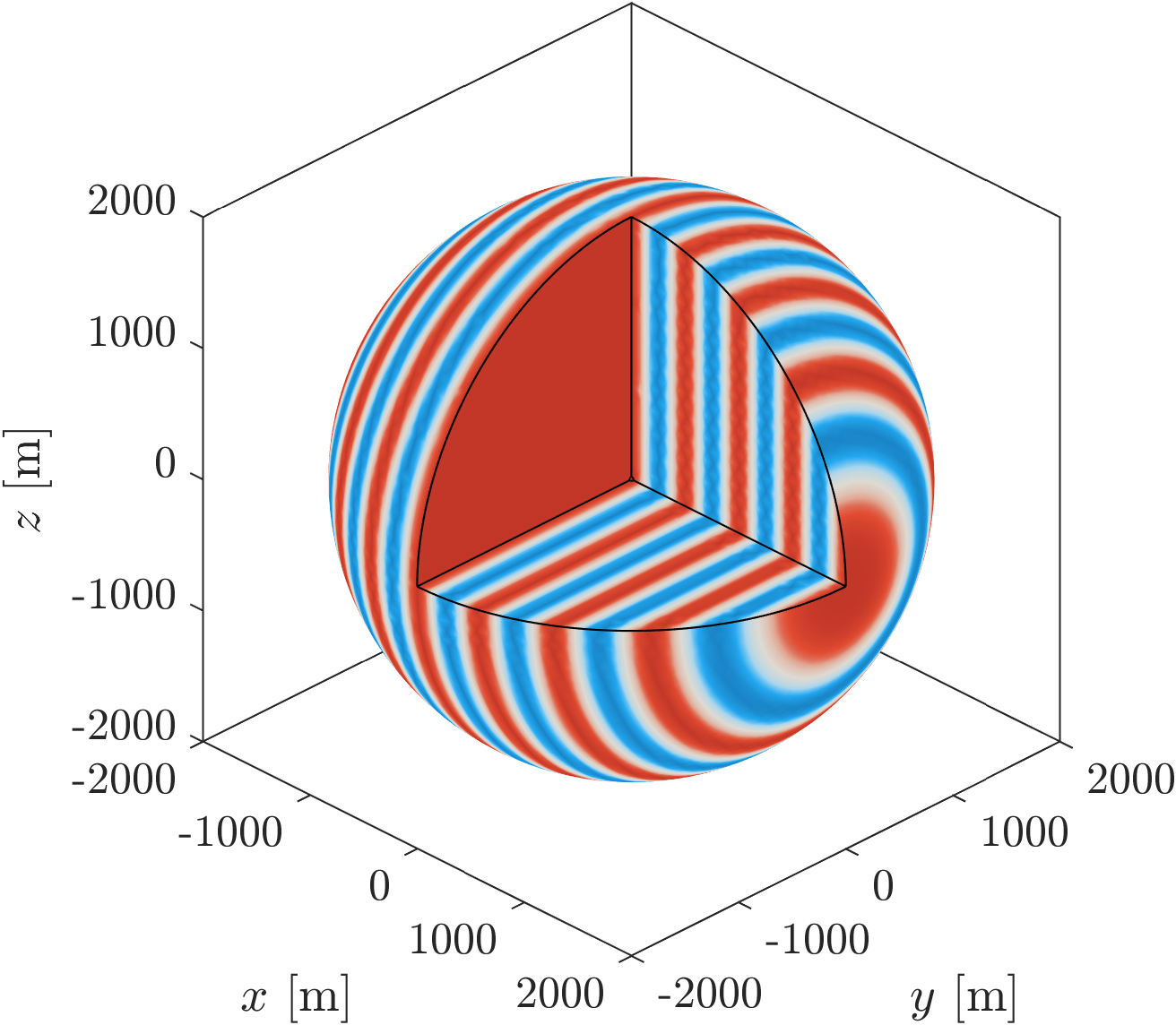} 
	\caption{(a) Finite element mesh of a sphere 
             with center at ${\bf x}_{0} = \{ 0,0,0 \}^{\mathrm{T}}$
             and radius $R=2000\,\mbox{m}$ 
             and a spherical cavity with radius $r_{0}=20\,\mbox{m}$,
             used to compute the Newtonian noise 
             due to a plane harmonic S-wave
             in a frequency range around 5\,Hz.
             (b) Displacement $\hat{u}_{z}({\bf x},\omega)$ 
             due to a plane harmonic S-wave 
             with unit amplitude and frequency of 5\,Hz
             propagating in the direction 
             ${\bf e}_{\mathrm{k}}=\{1,0,0\}^{\mathrm{T}}$
             with polarization vector
             ${\bf e}_{\mathrm{s}}=\{0,0,1\}^{\mathrm{T}}$.}
	\label{fig:fe-mesh-full-space-s}
\end{figure}
\par
For a plane harmonic S-wave propagating in a homogeneous linear elastic full space, 
an analytical expression for the Newtonian noise can be derived if wave scattering
by the cavity is ignored.
Inserting equation~\eqref{eq:us} into equation~\eqref{eq:nntf1} and evaluating 
the volume integral in spherical coordinates between radii $r_{0}$ and $R$ 
yields the following analytical expression for the Newtonian noise 
$\delta\hat{\bf a}_{\mathrm{t}}({\bf x}_{0},\omega)$ at the mirror position 
${\bf x}_{0}$~\cite{harm19a}:
\begin{eqnarray}
	\delta\hat{\bf a}_{\mathrm{t}}({\bf x}_{0},\omega) &=&
	-4 \pi \rho G
    \left(
    \frac{j_{1}(k_{\mathrm{s}} r_{0})}{k_{\mathrm{s}} r_{0}} -
    \frac{j_{1}(k_{\mathrm{s}} R)}{k_{\mathrm{s}} R}
    \right)
    \hat{\bf u}({\bf x}_{0}) \,,
    \label{eq:at-s-R}
\end{eqnarray}
which, for a fullspace ($R \rightarrow \infty$), gives:
\begin{eqnarray}
	\delta\hat{\bf a}_{\mathrm{t}}({\bf x}_{0}) &=&
	-4 \pi \rho G
    \frac{j_{1}(k_{\mathrm{s}} r_{0})}{k_{\mathrm{s}} r_{0}}
    \hat{\bf u}({\bf x}_{0}) \,.
	\label{eq:at-s-inf}
\end{eqnarray}
Since the divergence of the displacement field associated with an S-wave is 
equal to zero, an S-wave causes no bulk contribution to the Newtonian noise
and the total Newtonian noise is equal to the surface contribution.
\par
The Newtonian noise is computed using the same finite element mesh as for the
P-wave (figure \ref{fig:fe-mesh-full-space-s}a).
Figure~\ref{fig:verif-full-space-s}a compares the real part 
of the total Newtonian noise 
$\delta\hat{a}_{\mathrm{t}z}^{\mathrm{num}}({\bf x}_{0},\omega)$ 
computed with the finite element approximation with the real part of the analytical solution 
$\delta\hat{a}_{\mathrm{t}z}^{\mathrm{ref}}({\bf x}_{0},\omega)$ 
in the frequency range between 0.1\,Hz and 10\,Hz.
The displacement $\hat{\bf u}({\bf x}_{0},\omega)$ at the mirror in 
equation~(\ref{eq:at-s-inf}) is purely real,
so that $\delta\hat{a}_{\mathrm{t}z}({\bf x}_{0},\omega)$ is also real.
The imaginary part vanishes and therefore is not shown.
In the low frequency range considered, the total Newtonian noise 
$\delta\hat{a}_{\mathrm{t}z}^{\mathrm{ref}}({\bf x}_{0},\omega)$
is nearly constant.
The numerical prediction 
$\delta\hat{a}_{\mathrm{t}z}^{\mathrm{num}}({\bf x}_{0},\omega)$
agrees very well with the analytical result around 5\, Hz.
At this frequency, the extension of the finite element mesh is sufficiently large, while the finite element size is sufficiently small to capture the wavelength.
This is reflected by a low relative error $\varepsilon_{\mathrm{t}z}(\omega)$ in 
figure~\ref{fig:verif-full-space-s}b.
It is noted that the numerical solution converges faster to the analytical 
reference solution for an S-wave than for a P-wave. 
This is expected, since the domain size $R=2\lambda_{\mathrm{p}}^{\mathrm{max}}$ 
was tailored to the wavelength of the P-wave. The wavelength of the S-wave 
is two times smaller, so that $R=4\lambda_{\mathrm{s}}^{\mathrm{max}}$,
which reduces the truncation error.
At higher frequencies, however, the finite elements become  larger with respect to the S-wave wavelength, which results in a slight increase of the relative error.

\subsection{Harmonic Rayleigh wave in an elastic halfspace}
\label{subsec:verif-half-space-r}

\begin{figure}
   \centering
   \includegraphics[scale=0.7]{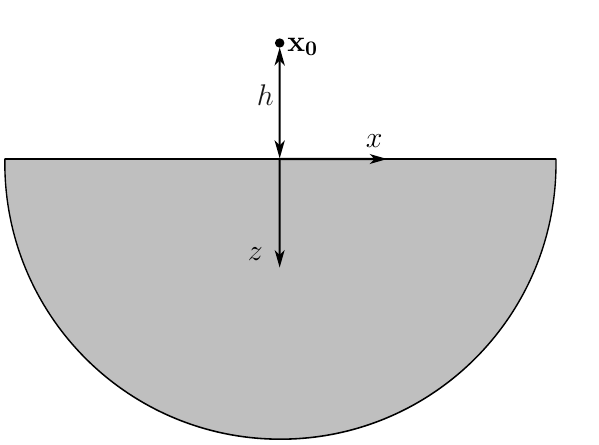}  
   \caption{Homogeneous halfspace with a test mass at a position  
            ${\bf x}_{0} = \left\{ 0,0,-h \right\}^{\mathrm{T}}$ above the free surface.}
   \label{fig:verif-halfspace}
\end{figure}

Next, the Newtonian noise on a test mass located at a position 
${\bf x}_{0} = \left\{ 0,0,-h \right\}^{\mathrm{T}}$ above the free surface of 
a homogeneous linear elastic halfspace (figure~\ref{fig:verif-halfspace}) due to a 
harmonic Rayleigh wave with a wavelength $\lambda_{\mathrm{R}}$ is considered.

\par
A Rayleigh wave propagates in a homogeneous linear elastic halfspace along the 
direction ${\bf k}_{\mathrm{k}} = k_{\mathrm{R}} {\bf e}_{\mathrm{k}}$, with 
$k_{\mathrm{R}} = \omega / C_{\mathrm{R}}$ the Rayleigh wavenumber, 
$C_{\mathrm{R}}$ the Rayleigh wave velocity, and 
${\bf e}_{\mathrm{k}} = \{\cos\theta,\sin\theta,0\}^{\mathrm{T}}$ 
a unit vector in a plane perpendicular to the $z$-axis.
The displacement vector $\hat{\bf u}({\bf x},\omega)$ can be written as \cite{rayl87a}:
\begin{eqnarray}
	\hat{\bf u}({\bf x},\omega) &=&
	\left(
    \hat{u}_{\mathrm{k}}(z,\omega) {\bf e}_{\mathrm{k}} + 
    \hat{u}_{z}(z,\omega) {\bf e}_{z}
    \right)
    \exp{(- \mathrm{i} {\bf k}_{\mathrm{k}} \cdot {\bf x})} \,,
    \label{eq:u_r}
\end{eqnarray}
where the horizontal component $\hat{u}_{\mathrm{k}}(z,\omega)$ along 
${\bf e}_{\mathrm{k}}$ is equal to:
\begin{eqnarray}
  \hat{u}_{\mathrm{k}}(z,\omega) &=&
   + \mathrm{i} k_{z\mathrm{s}} A_{\mathrm{s}} e^{-\mathrm{i} k_{z\mathrm{s}} z}
   - \mathrm{i} k_{\mathrm{R}}  A_{\mathrm{p}} e^{-\mathrm{i} k_{z\mathrm{p}} z} \,,
   \label{eq:rayleigh-uk}
\end{eqnarray}
while the vertical component $\hat{u}_{z}(z,\omega)$ is equal to:
\begin{eqnarray}
  \hat{u}_{z}(z,\omega) &=&
   - \mathrm{i} k_{\mathrm{R}}  A_{\mathrm{s}} e^{-\mathrm{i} k_{z\mathrm{s}} z}
   - \mathrm{i} k_{z\mathrm{p}} A_{\mathrm{p}} e^{-\mathrm{i} k_{z\mathrm{p}} z} \,.
   \label{eq:rayleigh-uz}
\end{eqnarray}
The vertical wavenumbers $k_{z \mathrm{p}}$ and $k_{z \mathrm{s}}$ follow from the 
dispersion relations:
\begin{eqnarray}
   k_{z \mathrm{p}} &=& 
   - \mathrm{i} \sqrt{k_{\mathrm{R}}^{2} - k_{\mathrm{p}}^{2}}\,, \\
   k_{z \mathrm{s}} &=& 
   - \mathrm{i} \sqrt{k_{\mathrm{R}}^{2} - k_{\mathrm{s}}^{2}} \,,
\end{eqnarray}
where $k_{\mathrm{p}}=\omega/C_{\mathrm{p}}$ and $k_{\mathrm{s}}=\omega/C_{\mathrm{s}}$ 
are the dilatational and shear wavenumbers.
Negative imaginary vertical wavenumbers are selected since the real solution 
$k_{\mathrm{R}}$ of Rayleigh's cubic characteristic equation \cite{rayl87a} is larger 
than both wavenumbers $k_{\mathrm{s}}$ and $k_{\mathrm{p}}$. 
When these vertical wavenumbers are inserted in the displacement components 
(\ref{eq:rayleigh-uk}) and (\ref{eq:rayleigh-uz}), an evanescent or surface wave
is obtained. 
The wave potentials $A_{\mathrm{p}}$ and $A_{\mathrm{s}}$ are components of the
corresponding eigenvector (the Rayleigh wave mode) and related through the 
traction-free boundary condition at the free surface $z=0$:
\begin{eqnarray}
  A_{\mathrm{s}} &=& \beta A_{\mathrm{p}} \,,
\end{eqnarray}
where the factor $\beta$ is defined as:
\begin{eqnarray}
  \beta &=& 
  \frac{2 k_{\mathrm{R}} k_{z \mathrm{p}}}{k_\mathrm{s}^{2} - 2 k_{\mathrm{R}}^{2}} \,.
\end{eqnarray}
The Rayleigh wave mode is subsequently scaled so that the vertical displacement 
at the surface $\hat{u}_{z}(0,\omega)$ is equal to 1:
\begin{eqnarray}
  \hat{u}_{z}(0,\omega) &=& 
  - \mathrm{i} \left( k_{\mathrm{R}} \beta + k_{z\mathrm{p}} \right) A_{\mathrm{p}}
  =  1 \,,
  \label{eq:uz0}
\end{eqnarray}
resulting in the following value for the potential $A_{\mathrm{p}}$:
\begin{eqnarray}
  A_{\mathrm{p}} &=& 
  \frac{\mathrm{i}}{k_{\mathrm{R}} \beta + k_{z\mathrm{p}}} \,.
  \label{eq:Ap_rayl}
\end{eqnarray}
\par
\begin{figure}[!htb]
	\centering
    (a) \includegraphics[width=0.425\textwidth]{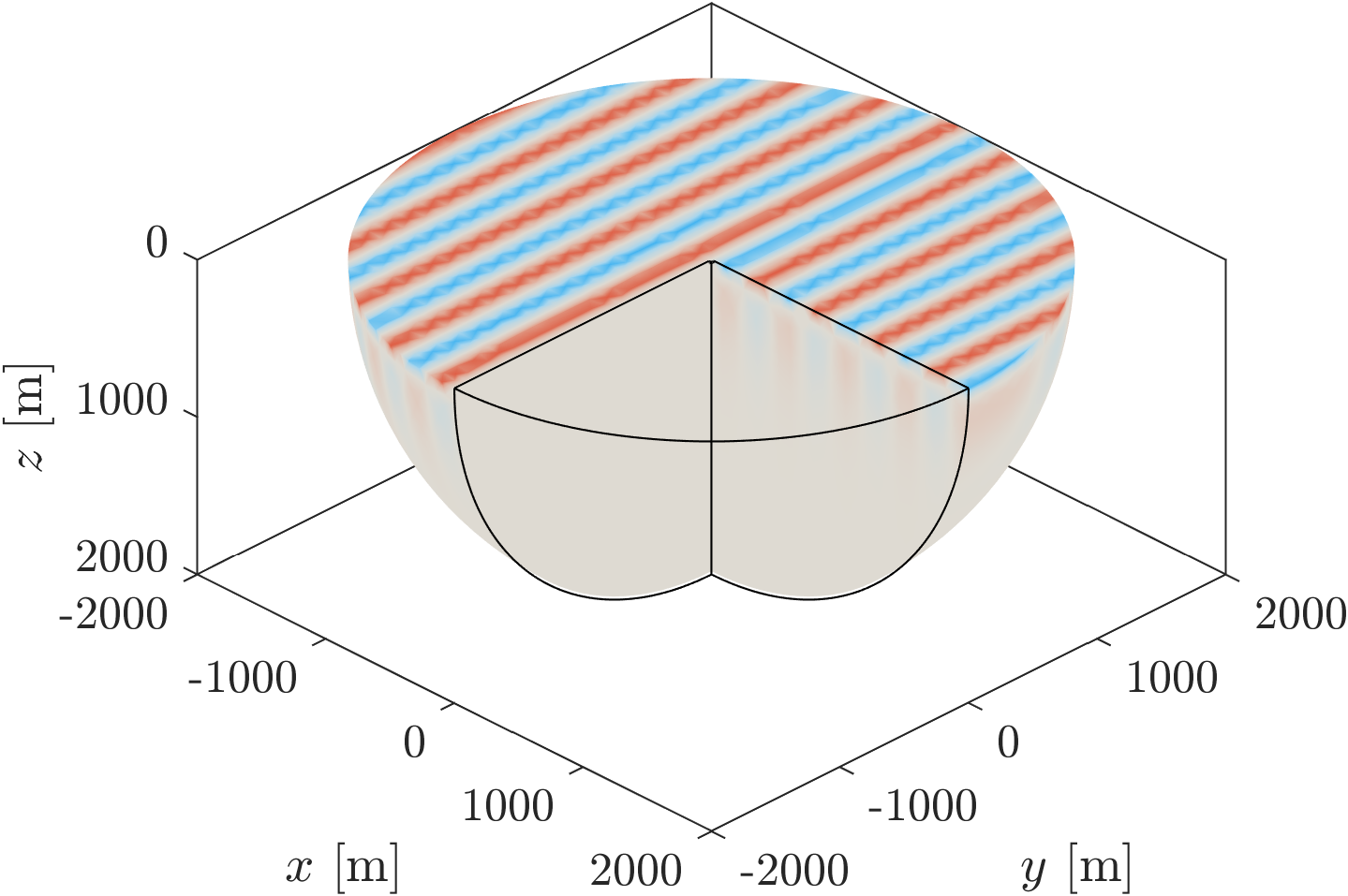} \quad
	(b) \includegraphics[width=0.425\textwidth]{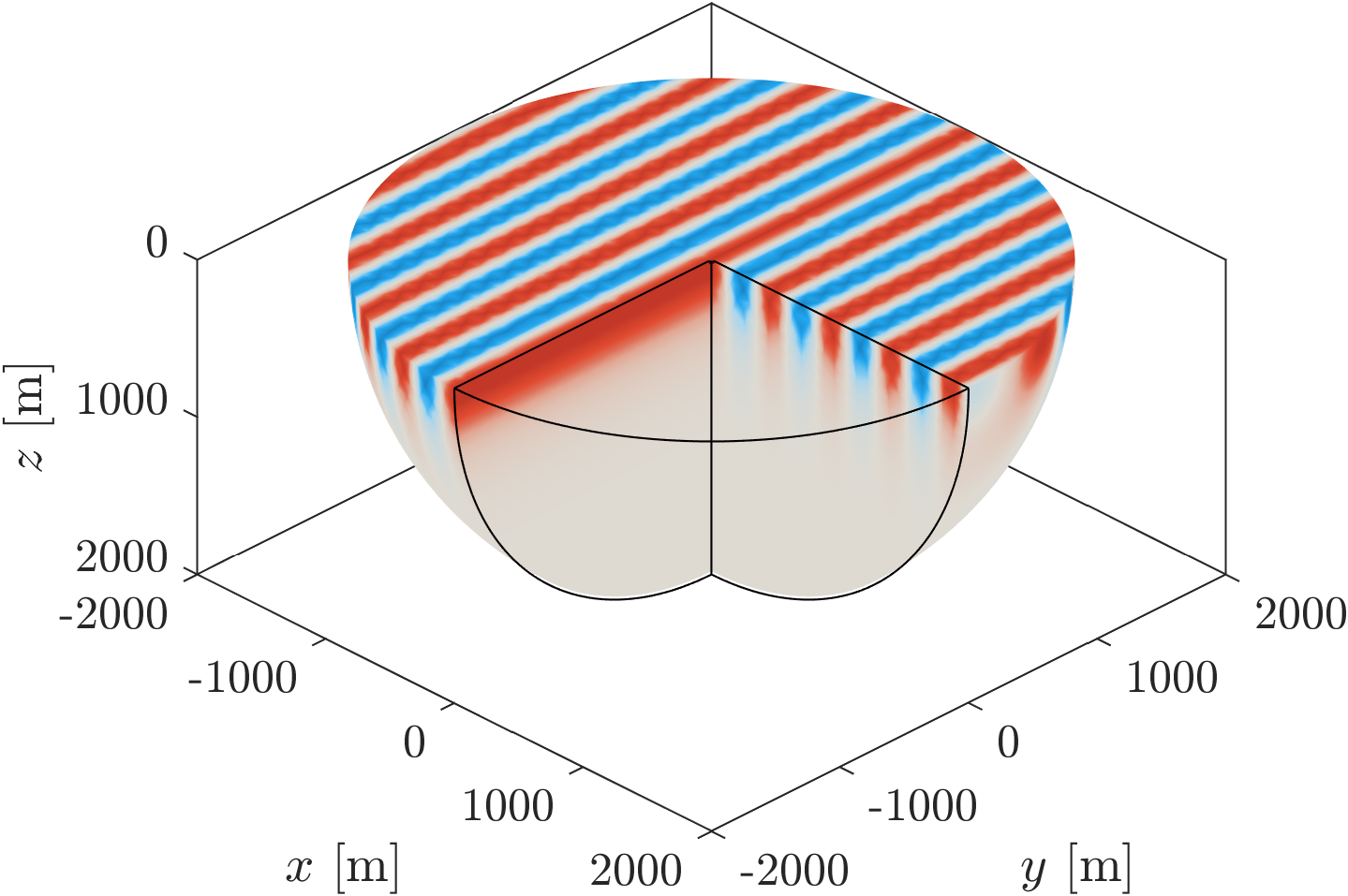} 
	\caption{Displacement 
             (a) $\hat{u}_{x}({\bf x},\omega)$ and 
             (b) $\hat{u}_{z}({\bf x},\omega)$
             due to a harmonic Rayleigh wave with unit amplitude
             and frequency of 5\,Hz
             propagating in the direction 
             ${\bf e}_{\mathrm{k}}=\{1,0,0\}^{\mathrm{T}}$.
             }
	\label{fig:verif-halfspace-r}
\end{figure}

Figure~\ref{fig:verif-halfspace-r} shows the displacements 
$\hat{u}_{x}({\bf x},\omega)$ and 
$\hat{u}_{z}({\bf x},\omega)$ of a harmonic Rayleigh wave with unit amplitude 
and frequency of 5\,Hz, propagating in the direction 
${\bf e}_{\mathrm{k}}=\{1,0,0\}^{\mathrm{T}}$ in a homogeneous linear elastic 
halfspace with the same properties as introduced for the full space. 
The wave field is shown on a finite element mesh that will subsequently be used to 
compute the Newtonian noise.
Both components exhibit exponential decay with depth.
The vertical displacement $\hat{u}_{z}({\bf x},\omega)$ has unit amplitude at 
the surface, while the horizontal displacement $\hat{u}_{x}({\bf x},\omega)$ 
changes sign around $z\approx0.2\lambda_{\mathrm{R}}$, which corresponds to 
the reversal of particle motion from retrograde near the surface to prograde 
at larger depth, characteristic of Rayleigh waves. 
\par
The analytical expression of the total Newtonian noise for a harmonic Rayleigh 
wave propagating in a homogeneous linear elastic halfspace and a mirror located at 
${\bf x}_{0} = \{ 0,0,-h \}^{\mathrm{T}}$ above the free surface is equal to
\cite{harm19a}:
\begin{eqnarray}
   \delta\hat{\bf a}_{\mathrm{t}}({\bf x}_{0},\omega) &=&
    2\pi G \rho \gamma \exp{(-k_{\mathrm{R}} h)}
    \left( \mathrm{i} {\bf e}_{\mathrm{k}} - {\bf e}_{z} \right) 
    \exp{(-\mathrm{i} {\bf k}_{\mathrm{k}}\cdot {\bf x}_{0})} \,,
    \label{eq:at_r}
\end{eqnarray}
where 
\begin{eqnarray}
    \gamma &=& 
    \frac{k_{\mathrm{R}}(1-\sqrt{k_{z\mathrm{p}}/k_{z\mathrm{s}}})}
         {\mathrm{i} k_{z\mathrm{p}} - k_{\mathrm{R}} 
         \sqrt{k_{z\mathrm{p}}/k_{z\mathrm{s}}}} \,.
\end{eqnarray}
\par
\begin{figure}[!htb]
   \centering
   \includegraphics[width=0.45\textwidth]{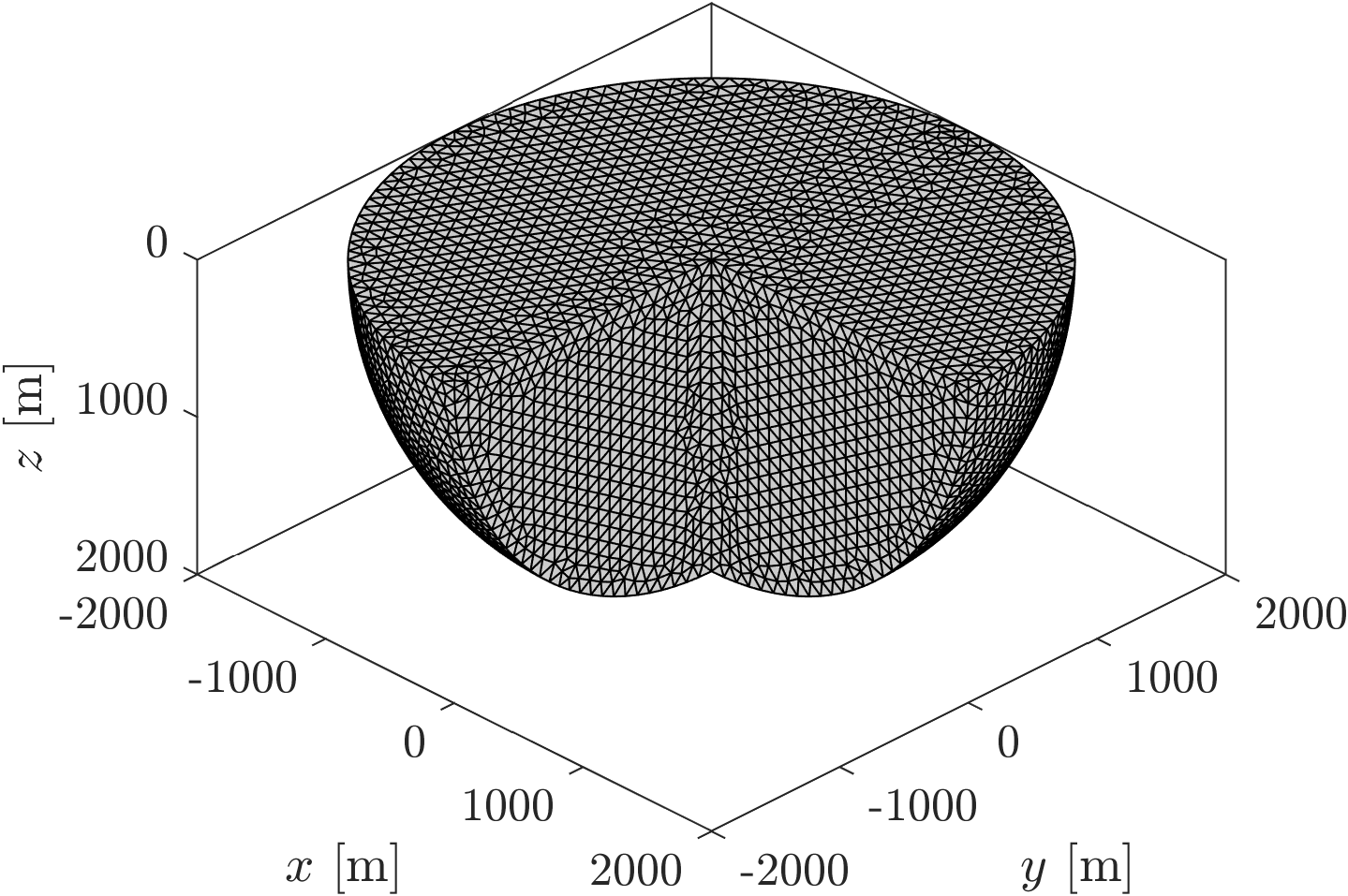} 
   \caption{Finite element mesh of a halfspace with radius $R=2000\,\mbox{m}$
            and centre at ${\bf x}_{0}=\{0,0,-20\}^{\mathrm{T}}$ 
            cut by the horizontal plane $z=0$, 
            used to compute the Newtonian noise due to a harmonic Rayleigh wave
            in a frequency range around 5\,Hz.
            }
	\label{fig:fe-mesh-halfspace}
\end{figure}

The finite element formulation is subsequently 
used to compute the Newtonian noise due to a harmonic Rayleigh wave on a mirror 
located at ${\bf x}_{0}=\{0,0,-20\}^{\mathrm{T}}$. 
A finite element mesh is created on a sphere with radius $R=2000 \,\mbox{m}$ and
center at ${\bf x}_{0}$, cut by the horizontal plane $z=0$ (figure~\ref{fig:fe-mesh-halfspace}). 
The same meshing criteria are applied as for the full space examples
in the previous subsections.
The minimum element size $l_{\mathrm{e}}$ 
is defined by the minimum
shear wavelength $\lambda_{\mathrm{s}}^{\mathrm{min}}/n$ with $n=5$.
\par
\begin{figure}[!htb]
	\centering
	(a) \includegraphics[width=0.425\textwidth]{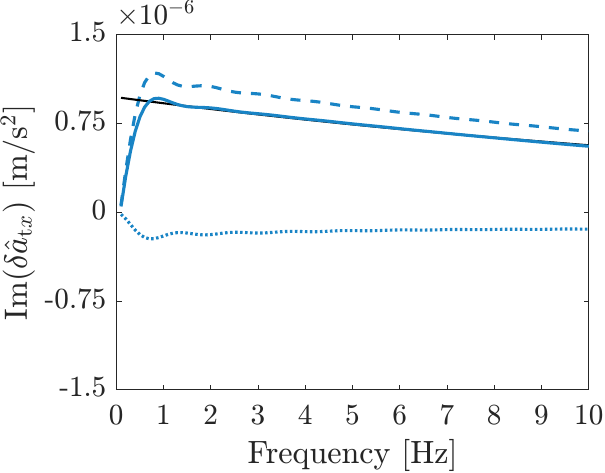} \quad
	(b) \includegraphics[width=0.425\textwidth]{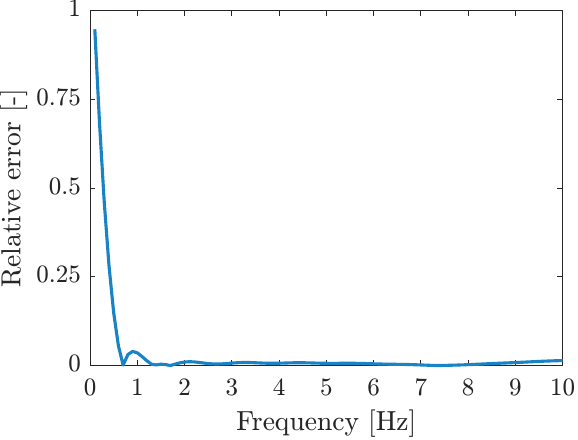}
    \caption{(a) Imaginary part of the total Newtonian noise 
             $\delta\hat{a}_{\mathrm{t}x}({\bf x}_{0},\omega)$ 
             computed analytically ($\textcolor{black}{\full}$) and 
             with the numerical model ($\textcolor{blue2}{\full}$), 
             due to a harmonic Rayleigh wave in a linear elastic halfspace
             with a test mass located at a height $h=20\,\mbox{m}$ above the free surface. 
             The bulk contribution 
             $\delta\hat{a}_{\mathrm{b}x}({\bf x}_{0},\omega)$ 
             ($\textcolor{blue2}{\dotted}$) and 
             the surface contribution 
             $\delta\hat{a}_{\mathrm{s}x}({\bf x}_{0},\omega)$ 
             ($\textcolor{blue2}{\dashed}$)
             are also shown.           
             (b) Relative error $\varepsilon_{\mathrm{t}x}(\omega)$ 
             on the total Newtonian noise.
             }
	\label{fig:verif-halfspace-ax}
\end{figure}

Figure~\ref{fig:verif-halfspace-ax}a compares the imaginary part of the total Newtonian 
noise $\delta\hat{a}_{\mathrm{t}x}^{\mathrm{num}}({\bf x}_{0},\omega)$ predicted with
the finite element formulation and the analytical reference solution in the frequency range 
between 0.1\,Hz and 10\,Hz.
Equation~(\ref{eq:at_r}) shows that the horizontal component $\delta\hat{a}_{\mathrm{t}x}$
is purely imaginary. The real part vanishes and therefore is not shown.
At low frequencies, the numerical result deviates from the reference solution due to 
the finite radius $R$ of the domain, for similar reasons as discussed for the full space cases.
At frequencies around 5\,Hz, excellent agreement is observed.
At higher frequencies, a slight increase
of the relative error is observed as the wavelength decreases with respect to the adopted finite element size.
This is reflected by the relative error 
$\varepsilon_{\mathrm{t}x}(\omega)$ in figure~\ref{fig:verif-halfspace-ax}b,
confirming that the adopted framework remains accurate for the halfspace configuration. 
The total Newtonian noise computed from equation~\eqref{eq:fe-at} is, as expected,
equal to the sum of the bulk and surface contributions obtained from 
equations~\eqref{eq:fe-ab} and~\eqref{eq:fe-as}.
\par
\begin{figure}[!htb]
	\centering
	(a) \includegraphics[width=0.425\textwidth]{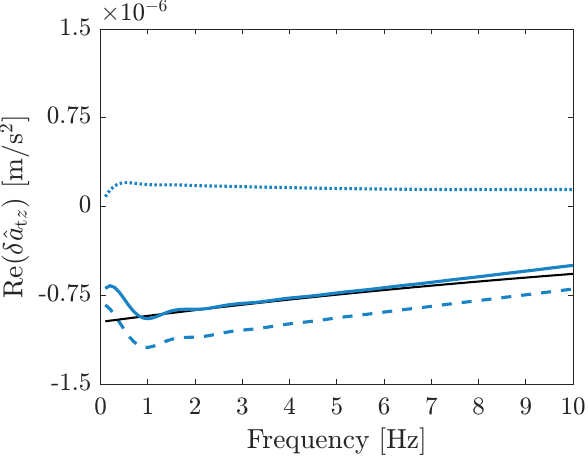} \quad
	(b) \includegraphics[width=0.425\textwidth]{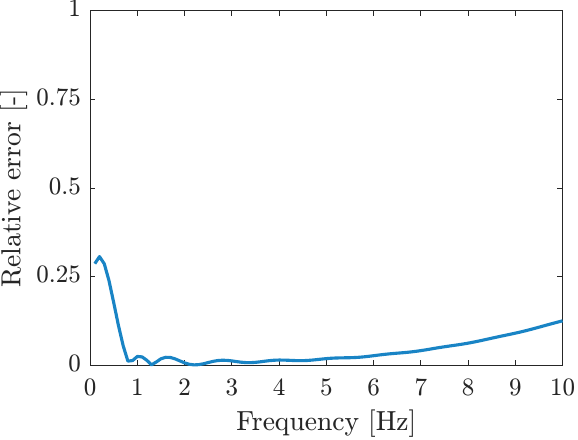}
    \caption{(a) Real part of the total Newtonian noise 
             $\delta\hat{a}_{\mathrm{t}z}({\bf x}_{0},\omega)$ 
             computed analytically ($\textcolor{black}{\full}$) and 
             with the numerical model ($\textcolor{blue2}{\full}$), 
             due to a harmonic Rayleigh wave in a linear elastic halfspace
             with a test mass located at a height $h=20\,\mbox{m}$ above the free surface. 
             The bulk contribution 
             $\delta\hat{a}_{\mathrm{b}z}({\bf x}_{0},\omega)$ 
             ($\textcolor{blue2}{\dotted}$) and 
             the surface contribution 
             $\delta\hat{a}_{\mathrm{s}z}({\bf x}_{0},\omega)$ 
             ($\textcolor{blue2}{\dashed}$)
             are also shown.           
             (b) Relative error $\varepsilon_{\mathrm{t}z}(\omega)$ 
             on the total Newtonian noise.
             }
	\label{fig:verif-halfspace-az}
\end{figure}

Figure~\ref{fig:verif-halfspace-az}a compares the real part of the total Newtonian noise
$\delta\hat{a}_{\mathrm{t}z}^{\mathrm{num}}({\bf x}_{0},\omega)$ 
predicted with the finite element formulation and the analytical reference solution
in the frequency range between 0.1\,Hz and 10\,Hz.
According to equation~(\ref{eq:at_r}), the vertical component $\delta\hat{a}_{\mathrm{t}z}$ 
is real. The imaginary part vanishes and therefore is not shown.
As for the horizontal component, the low-frequency response is affected by the finite
radius $R$ of the domain. At frequencies around 5\,Hz, excellent agreement is observed.
At higher frequencies, a larger increase of the relative error is observed as the
wavelength decreases with respect to the adopted finite element size, as reflected by
the relative error $\varepsilon_{\mathrm{t}z}(\omega)$ in
figure~\ref{fig:verif-halfspace-az}b.

\section{Conclusions}
\label{sec:conclusion}

This paper presents an efficient finite element formulation for 
computing Newtonian noise on a mirror in a cavity from a 3D seismic wave field. Gaussian quadrature on a 3D finite element mesh is used to numerically 
evaluate the volume and surface integrals that define the total Newtonian noise,
as well as the bulk and surface contributions. 
Since the finite element matrices depend only on the geometry and density of the
problem domain, they are computed once and subsequently used for arbitrary seismic 
wave fields, yielding an efficient evaluation of Newtonian noise through a 
matrix–vector multiplication.
\par
The formulation is implemented in the ANNA Newtonian Noise Analysis toolbox in 
the MATLAB programming and numeric computing platform and is compatible with the 
open-source GNU Octave Scientific Programming Language and a Python version is also available.
\par
Newtonian noise computed with ANNA is verified for a suite of problems for which 
an analytical solution is available.
Newtonian noise is predicted for a plane harmonic P- and S-wave propagating in
a homogeneous linear elastic full space with a cavity with radius $r_{0}$ and for
a test mass at a distance $h$ above the free surface of a homogeneous linear 
elastic halfspace where a harmonic Rayleigh wave propagates along the surface.
When \textendash~in the frequency range of interest \textendash~ proper criteria 
for finite element mesh extension are met carefully, the predicted 
Newtonian noise is in very good agreement with the reference solutions.
Therefore, ANNA can be employed with confidence 
for computationally efficient Newtonian noise predictions from 
wave fields originating from different anthropogenic sources on soil domains with
a more complex topography and stratification. 
The present verification can then be followed by a validation based on recorded 
seismic wave fields.

\section*{Acknowledgments}

Results presented in this paper have been obtained within the frame of the 
FWO-IRI projects I002123N "Essential Technologies for the Einstein Telescope" (2023-2024)
and I000725N "ET-TECH: Empowering Tomorrow's Technological Horizons for Einstein Telescope" (2025)
and the FWO Research Collaboration "Modelling external vibration sources" (2026) with the ET-EMR Project Office.
The financial support of the Research Foundation Flanders (FWO) is gratefully
acknowledged.

\section*{Data availability statement}

The MATLAB and Python implementations, together with a user manual can be downloaded from 
https://bwk.kuleuven.be/bwm/anna.

\appendix
\renewcommand\thefigure{A.\arabic{figure}}  
\section{Reynolds' transport theorem}
\label{app:rtt}
\setcounter{figure}{0}    

\begin{figure}[h!]
	\centering
	\includegraphics[scale=1]{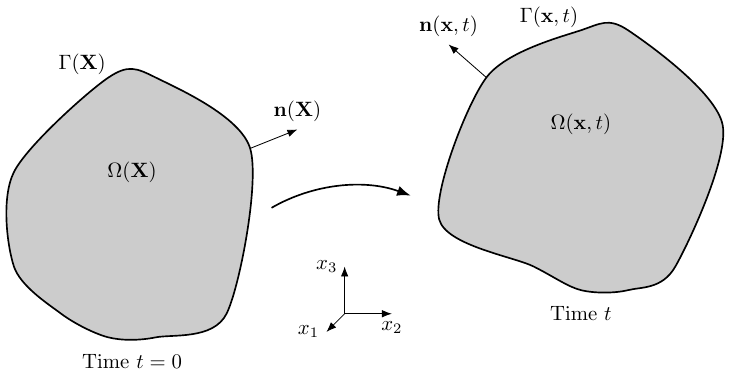}
	\caption{Material volume $\Omega({\bf X})$ 
    with boundary $\Gamma({\bf X})$ and 
    unit outward normal vector ${\bf n}({\bf X})$
    in the reference configuration at time $t=0$
	and material volume $\Omega({\bf x},t)$ 
    with boundary $\Gamma({\bf x},t)$ and 
    unit outward normal vector ${\bf n}({\bf x},t)$
    in the actual configuration at time $t$.}
	\label{fig:rtt1}
\end{figure}

Assume a set of particles that occupy at time $t=0$ (in the reference
configuration) a volume $\Omega({\bf X})$ with boundary $\Gamma({\bf X})$
and unit outward normal vector ${\bf n}({\bf X})$.
After deformation, at time $t$ (in the actual configuration), these
particles occupy a volume $\Omega({\bf x},t)$ with boundary 
$\Gamma({\bf x},t)$ and unit outward normal vector ${\bf n}({\bf x},t)$
(figure \ref{fig:rtt1}).
\par
We consider a mass dependent scalar or vector quantity $f(t)$, defined in the 
actual configuration as the volume integral of a continuously differentiable 
scalar or vector quantity $g({\bf x},t)$:
\begin{eqnarray}
   f(t) &=&
   \int_{\Omega({\bf x},t)} g({\bf x},t) 
   \, \mathrm{d}v \,.
   \label{eq:rtt1}
\end{eqnarray}
The scalar or vector quantity $g({\bf x},t)$ can be defined, for
example, as:
\begin{itemize}
\item the density $\rho({\bf x},t)$, in which case $f(t)$ represents 
the mass of the particles contained in the volume $\Omega({\bf x},t)$;
\item the quantity $\rho({\bf x},t) {\bf v}({\bf x},t)$, such that the 
vector $f(t)$ represents the linear momentum of the particles contained 
in the volume $\Omega({\bf x},t)$;
\item the quantity $- G \rho({\bf x},t) / \nrm{ {\bf x} - {\bf x}_{0} }$,
such that $f(t)$ represents the gravitational potential of the test 
mass at position ${\bf x}_{0}$.
\end{itemize}
\par
\begin{figure}[!htb]
	\centering
	\includegraphics[scale=1]{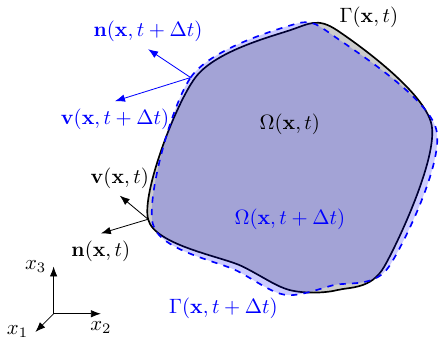}
	\caption{Material volume $\Omega({\bf x},t)$ with boundary 
    $\Gamma({\bf x},t)$ in the actual configuration at time $t$ and
    material volume $\Omega({\bf x},t+\Delta t)$ 
    with boundary $\Gamma({\bf x},t+\Delta t)$ 
    in the actual configuration at time $t+\Delta t$.}
	\label{fig:rtt2}
\end{figure}

We need to evaluate the following total derivative with respect to time,
as a rigorous representation of a "variation" of the quantity $f(t)$ 
(in order to express conservation of mass, conservation of linear momentum, 
or a variation of the gravitational potential):
\begin{eqnarray}
   \frac{D f(t)}{Dt} &=&
   \frac{D}{Dt} 
   \int_{\Omega({\bf x},t)} g({\bf x},t) 
   \, \mathrm{d}v \,.
   \label{eq:rtt2}
\end{eqnarray}
Reynolds' transport theorem gives a formal expression of the total
or material derivative with respect to time $t$ of the quantity $f(t)$,
or the variation of this quantity observed when following a set
of particles that were contained in a volume $\Omega({\bf X})$ at 
time $t=0$ and are contained in a volume $\Omega({\bf x},t)$ at time $t$
\cite{papa20a}:
\begin{eqnarray}
   \frac{D f(t)}{Dt} &=&
   \int_{\Omega({\bf x},t)}
   \left[ 
   \frac{D g({\bf x},t)}{D t} +
   g({\bf x},t) \nabla \cdot {\bf v}({\bf x},t)
   \right] 
   \, \mathrm{d}v
   \label{eq:rtt3a}
   \\ &=&
   \int_{\Omega({\bf x},t)}
   \left[ 
   \frac{\partial g({\bf x},t)}{\partial t} +
   {\bf v}({\bf x},t) \cdot \nabla g({\bf x},t) +
   g({\bf x},t) \nabla \cdot {\bf v}({\bf x},t)
   \right] 
   \, \mathrm{d}v
   \label{eq:rtt3b}
   \\ &=&     
   \int_{\Omega({\bf x},t)}
   \left[ 
   \frac{\partial g({\bf x},t)}{\partial t} +
   \nabla \cdot ( g({\bf x},t) {\bf v}({\bf x},t) ) 
   \right] 
   \, \mathrm{d}v
   \label{eq:rtt3c}
   \\ &=&     
   \int_{\Omega({\bf x},t)}
   \frac{\partial g({\bf x},t)}{\partial t} 
   \, \mathrm{d}v +
   \int_{\Gamma({\bf x},t)}
   g({\bf x},t) {\bf v}({\bf x},t) \cdot {\bf n}({\bf x},t)
   \, \mathrm{d}a \,.
   \label{eq:rtt3d}
\end{eqnarray}
The first term on the right-hand side of equation (\ref{eq:rtt3d}) is 
the instantaneous variation of the quantity $g({\bf x},t)$ or
the rate of change of $g({\bf x},t)$ at time $t$ 
for all particles contained in the volume $\Omega({\bf x},t)$.
The second term is the flux of the property $g({\bf x},t)$ as particles exit
the volume $\Omega({\bf x},t)$ through its surface $\Gamma({\bf x},t)$ with
a normal velocity ${\bf v}({\bf x},t) \cdot {\bf n}({\bf x},t)$, where
${\bf n}({\bf x},t)$ is the unit outward normal vector to the boundary 
$\Gamma({\bf x},t)$.
This term originates from the fact that the volume $\Omega({\bf x},t)$ with
boundary $\Gamma({\bf x},t)$ that contains a fixed amount of material has 
moved during a time step $\Delta t$ to a volume $\Omega({\bf x},t+\Delta t)$ 
with boundary $\Gamma({\bf x},t+\Delta t)$ (figure \ref{fig:rtt2}).
\par
In the particular case where the volume $\bar{\Omega}$ is a fixed
region in space (or a control volume) with boundary $\bar{\Gamma}$,
the integration over the volume is decoupled from partial integration
and Reynolds' transport theorem can be expressed as \cite{papa20a}:
\begin{eqnarray}
   \frac{D f(t)}{Dt} &=&
   \frac{\partial}{\partial t} 
   \int_{\bar{\Omega}}
   g({\bf x},t) 
   \, \mathrm{d}v +
   \int_{\bar{\Gamma}}
   g({\bf x},t) {\bf v}({\bf x},t) \cdot {\bf n}({\bf x},t)
   \, \mathrm{d}a \,.
   \label{eq:rtt4}
\end{eqnarray}
The first term on the right-hand side of equation (\ref{eq:rtt4}) is the
rate of change of the integral of the quantity $g({\bf x},t)$ over the 
volume $\bar{\Omega}$ due to its explicit dependence on time.
The second term is the flux of the property $g({\bf x},t)$ as particles exit
the volume $\bar{\Omega}$ through its fixed boundary $\bar{\Gamma}$ with
a normal velocity ${\bf v}({\bf x},t) \cdot {\bf n}({\bf x},t)$.

\section{Conservation of mass}
\label{app:com}

The total mass $M(t)$ of all particles contained at time $t$ in a volume
$\Omega({\bf x},t)$ is equal to:
\begin{eqnarray}
   M(t) &=& 
   \int_{\Omega({\bf x},t)} \rho({\bf x},t) 
   \, \mathrm{d}v \,,
   \label{eq:com1}
\end{eqnarray}
where $\rho({\bf x},t)$ is the density of the material.
\par
Conservation of mass is expressed as:
\begin{eqnarray}
   \frac{D M(t)}{Dt} &=& 
   \frac{D}{Dt} \int_{\Omega({\bf x},t)} \rho({\bf x},t) 
   \, \mathrm{d}v
   \, = \,
   0 
   \label{eq:com2}
\end{eqnarray}
and elaborated in one of the following forms applying Reynolds' transport theorem:
\begin{eqnarray}
   \lefteqn{
   \int_{\Omega({\bf x},t)}
   \left[ 
   \frac{D \rho({\bf x},t)}{D t} +
   \rho({\bf x},t) \nabla \cdot {\bf v}({\bf x},t)
   \right] 
   \, \mathrm{d}v
   \, = \, 0
   }
   \label{eq:com3a}
   \\ & \Leftrightarrow &
   \int_{\Omega({\bf x},t)}
   \left[ 
   \frac{\partial \rho({\bf x},t)}{\partial t} +
   {\bf v}({\bf x},t) \cdot \nabla \rho({\bf x},t) +
   \rho({\bf x},t) \nabla \cdot {\bf v}({\bf x},t)
   \right] 
   \, \mathrm{d}v
   \, = \, 0
   \label{eq:com3b}
   \\ & \Leftrightarrow &
   \int_{\Omega({\bf x},t)}
   \left[ 
   \frac{\partial \rho({\bf x},t)}{\partial t} +
   \nabla \cdot \left( \rho({\bf x},t) {\bf v}({\bf x},t) \right) 
   \right] 
   \, \mathrm{d}v
   \, = \, 0
   \label{eq:com3c}
   \\ & \Leftrightarrow &
   \int_{\Omega({\bf x},t)}
   \frac{\partial \rho({\bf x},t)}{\partial t} 
   \, \mathrm{d}v +
   \int_{\Gamma({\bf x},t)}
   \rho({\bf x},t) {\bf v}({\bf x},t) \cdot {\bf n}({\bf x},t)
   \, \mathrm{d}a
   \, = \, 0 \,.
   \label{eq:com3d}
\end{eqnarray}
In the special case where the volume $\bar{\Omega}$ is a fixed
region in space (or a control volume) with boundary $\bar{\Gamma}$,
the integration over the volume is decoupled from partial integration 
\cite{papa20a}, and conservation of mass reads as:
\begin{eqnarray}
   \frac{\partial}{\partial t} 
   \int_{\bar{\Omega}}
   \rho({\bf x},t) 
   \, \mathrm{d}v +
   \int_{\bar{\Gamma}}
   \rho({\bf x},t) {\bf v}({\bf x},t) \cdot {\bf n}({\bf x},t)
   \, \mathrm{d}a
   &=& 0 \,.
   \label{eq:com4}
\end{eqnarray}
\par
Equation (\ref{eq:com4}) expresses that the variation of mass in
the volume $\bar{\Omega}$ as a function of time (the first term)
is equal to minus the flux of mass through the boundary $\bar{\Gamma}$
(the second term) with a normal velocity
${\bf v}({\bf x},t) \cdot {\bf n}({\bf x},t)$.
\par
As conservation of mass must hold for any volume $\Omega({\bf x},t)$, 
the integrand in equations (\ref{eq:com3a}-\ref{eq:com3c}) must also
be equal to zero:
\begin{eqnarray}
   \lefteqn{
   \frac{D \rho({\bf x},t)}{D t} +
   \rho({\bf x},t) \nabla \cdot {\bf v}({\bf x},t)
   \, = \, 0
   }
   \label{eq:com5a}
   \\ & \Leftrightarrow &
   \frac{\partial \rho({\bf x},t)}{\partial t} +
   {\bf v}({\bf x},t) \cdot \nabla \rho({\bf x},t) +
   \rho({\bf x},t) \nabla \cdot {\bf v}({\bf x},t)
   \, = \, 0
   \label{eq:com5b}
   \\ & \Leftrightarrow &
   \frac{\partial \rho({\bf x},t)}{\partial t} +
   \nabla \cdot \left( \rho({\bf x},t) {\bf v}({\bf x},t) \right)
   \, = \, 0 \,.
   \label{eq:com5c}
\end{eqnarray}
In a rate-independent form where the total derivative of the mass 
potential with respect to time is replaced by a variation, and the 
velocity vector is replaced by the displacement vector, the variation 
of the density becomes:
\begin{eqnarray}
   \delta \rho({\bf x},t) +
   \nabla \cdot \left( \rho({\bf x},t) {\bf u}({\bf x},t) \right)
   &=& 0 \,.
   \label{eq:com6}
\end{eqnarray}

\bibliographystyle{unsrt}
\bibliography{abbrev,paper}

\end{document}